\documentclass[aps,prd,showpacs,nobibnotes,twocolumn,preprintnumbers,nobalancelastpage,superscriptaddress]{revtex4}
\usepackage{graphicx}
\usepackage{latexsym,hyperref}
\usepackage{dcolumn}
\usepackage{bm}
\usepackage{natbib}
\usepackage{color}

\input{epsf}
\newcommand{\beq}{\begin{equation}}
\newcommand{\eeq}{\end{equation}}
\newcommand{\bea}{\begin{eqnarray}}
\newcommand{\eea}{\end{eqnarray}}
\newcommand{\bi}{\begin{itemize}}
\newcommand{\ei}{\end{itemize}}
\newcommand{\bfi}{\begin{figure}[!t]
\epsfxsize=7cm
\epsffile}
\newcommand{\bfib}{\begin{figure}[htb]
\epsfxsize=9cm
\epsffile}
\newcommand{\bfig}{\begin{figure*}[htb]
\epsfxsize=12cm
\epsffile}
\newcommand{\efi}{\end{figure}}
\newcommand{\efib}{\end{figure}}
\newcommand{\efig}{\end{figure*}}

\newcommand{\la}{\lesssim}
\newcommand{\ga}{\gtrsim}
\newcommand{\aj}{AJ}
\newcommand{\apjl}{ApJL}
\newcommand{\mnras}{MNRAS}
\newcommand{\jcap}{JCAP}
\newcommand{\physrep}{Physics reports}

\newcommand{\bfs}{\mbox{\boldmath$s$}}
\newcommand{\bfr}{\mbox{\boldmath$r$}}
\newcommand{\bfx}{\mbox{\boldmath$x$}}
\newcommand{\bfk}{\mbox{\boldmath$k$}}
\newcommand{\bfv}{\mbox{\boldmath$v$}}
\newcommand{\bfu}{\mbox{\boldmath$u$}}
\newcommand{\bfp}{\mbox{\boldmath$p$}}
\newcommand{\bfq}{\mbox{\boldmath$q$}}
\newcommand{\Pdd}{P_{\delta\delta}}
\newcommand{\Pdt}{P_{\delta \Theta}}
\newcommand{\Ptt}{P_{\Theta\Theta}}

\newcommand{\hompc}{\,h\,{\rm Mpc}^{-1}}

\newcommand{\ompc}{\,{\rm Mpc}^{-1}}
\newcommand{\mpcoh}{\,h^{-1}\,{\rm Mpc}}

\begin{document}
\title{Study on the mapping of dark matter clustering from real space to redshift space}
\author{Yi Zheng}
\email[Email at ]{yizheng@kasi.re.kr}
\affiliation{Korea Astronomy and Space Science Institute, Daejeon 34055, Republic of Korea}
\author{Yong-Seon Song}
\email[Email at ]{ysong@kasi.re.kr}
\affiliation{Korea Astronomy and Space Science Institute, Daejeon 34055, Republic of Korea}

\begin{abstract}
The mapping of dark matter clustering from real space to redshift space introduces the anisotropic property to the measured density power spectrum in redshift space, known as the redshift space distortion effect. The mapping formula is intrinsically non-linear, which is complicated by the higher order polynomials due to indefinite cross correlations between the density and velocity fields, and the Finger--of--God effect due to the randomness of the peculiar velocity field. Whilst the full higher order polynomials remain unknown, the other systematics can be controlled consistently within the same order truncation in the expansion of the mapping formula, as shown in this paper. The systematic due to the unknown non--linear density and velocity fields is removed by separately measuring all terms in the expansion directly using simulations. The uncertainty caused by the velocity randomness is controlled by splitting the FoG term into two pieces, 1) the ``one--point" FoG term being independent of the separation vector between two different points, and 2) the ``correlated" FoG term appearing as an indefinite polynomials which is expanded in the same order as all other perturbative polynomials. Using 100 realizations of simulations, we find that the Gaussian FoG function with only one scale--independent free parameter works quite well, and that our new mapping formulation accurately reproduces the observed 2--dimensional density power spectrum in redshift space at the smallest scales by far, up to $k\sim 0.2\hompc$, considering the resolution of future experiments.
\end{abstract}
\maketitle

\section{Introduction}
\label{sec:Intro}

The presence of the cosmic acceleration has been confirmed by multiple experiments since 1998 \cite{Acceleration1,Acceleration2}, supporting the standard model of the universe, which is dominated by dark materials motivating new physics \cite{weinbergreview}. There has been significant theoretical and observational studies exploring the true nature of the cosmic acceleration. Some predict the unknown materials such as the dark energy expelling the cosmic expansion against the gravitational force \cite{Amendola05,Yamamoto05,Wang08,Percival09,Song09,White09,Song10,Wang10},      and others propose the modified gravity which is the consequence of our incomplete knowledge of the gravitational theory \cite{Zhang07c,Jain08,Linder08,Reyes10,Cai12,Gaztanaga12,Jennings12,Li12,Okumura15}. Whichever it might be true, the outcome will revolutionize our understanding of the fundamental physics.

The nature of the dark universe can be revealed by observing the large scale structure of the universe. Combined with the galaxy angular positions, the redshifts of galaxies at a selected sky area are measured to reveal the 3D distribution of the galaxies. The large scale structure of the matter can be estimated from this observed redshift map, which provides us an opportunity to probe in precision the key observables  demanded for explaining the dark materials of the universe. First, the galaxy clustering at large scales contains the fossil information of the primordial baryonic plasma from the last scattering epoch, called the baryon acoustic oscillations (hereafter BAO). The characteristic BAO scale determined by the cosmic microwave background (hereafter CMB) experiments offers us a standard ruler which allows us to measure the distances to the selected galaxies with a high accuracy \cite{Seo03,Eisenstein05,Blake11a,Anderson12,
Kazin14,Song14,Anderson14,Gilmarn15,Zhao16}. Second, the observed clustering pattern of galaxy power spectrum becomes anisotropic with the presence of the apparent mismatch between the underlying theoretical model and the true universe, as the redshift and angular positions of galaxies need to be converted to the co--moving radial and transverse distances using theoretical model. This is known as the Alcock-Paczynski (A-P) effect \cite{APtest}. Third, the observed galaxy clustering provides us an unique window accessing  the evolution history of the cosmic structure formation. The observed galaxy clustering seen by the spectroscopic measurements is distorted along the line of sight by the peculiar velocity of galaxies, known as the redshift space distortions (hereafter RSD) effect \cite{Jackson72,Sargent77,Peebles80,Kaiser87,Peacock94,Ballinger96}. The strength of the anisotropic pattern imprinted on the power spectrum can be used to determine the linear growth rate of the structure formation \cite{Peacock01,Tegmark02,Tegmark04,Samushia12,Guzzo08,Blake11b,Blake12,Reid12,Tojeiro12,
Reid14,Song15a,Song15b,Okumura15,Zhao16,Simpson16}.

However, the full statistical analysis of the anisotropy pattern of the RSD effect is contaminated by several systematics \cite{Peebles80,Fisher95,Heavens98,White01,Seljak01,Kang02,Tinker06,Tinker07,
Scoccimarro04,Matsubara08a,Matsubara08b,Desjacques10,Taruya10,Taruya13,
Matsubara11,Okumura11a,Okumura11b,Sato11,Jennings11b,
Reid11,Seljak11,Okumura12,Okumura12b,Kwan12,Zhangrsd,Zheng13,Ishikawa14,White15,Jennings16,Bianchi15,Bianchi16,Simpson16}, which weakens our confidence on RSD cosmological constraints in comparison with other probes of large scale structure \cite{Okumura11a,Jennings11a,Jennings11b,Kwan12,Bianchi12,delaTorre12,White15}. First, the transformation between real and redshift spaces is intrinsically non--linear, in that the density perturbation and velocity fields are non--linearly coupled together in the mapping. The factorized formula has been proposed to achieve the RSD theoretical model, which turns out to be the combination of non--separable linear squeezing effect and non--linear smearing effects caused by those higher order polynomials. Next, the non--linear corrections on density and velocity fields are poorly understood and hard to be cleaned from the measured anisotropic power spectrum. Several perturbation theory with different truncation strategies, e.g., the closure approximation \cite{Closure}, have been exploited to predict the non--linear density--density, density--velocity and velocity--velocity power spectra, but all of them are not trustable beyond the limit of approximation. Finally, the small scale distortion along the line of sight is caused by the randomness of the velocity field, called the Finger--of--God (hereafter FoG) effect. This effect is intrinsically non--perturbative, and there is little clue of its exact theoretical form neither in scale nor in time. All these systematics are mixed up in the full RSD analysis, which makes us losing our confidence on cosmological constraints from the RSD effect beyond the conservative limited scale at quite linear regime.

In this paper, we study solutions to remove the above systematics and to reproduce the observed power spectrum in accuracy. The mapping formula of clustering from real to redshift space is given in an unbiased way. The full higher order polynomials are given by infinite cross correlation terms between density and velocity fields, which will not be provided in any closed form but simply expanded and truncated at certain order. Meanwhile, other systematics can be removed at least in the same expansion order of polynomials. In the expansion of higher order polynomials, there are multi--point power spectra whose non--linear corrections are unknown. Theoretical non--linear approximation can be used, but it is difficult to theoretically formulate higher order correlation functions in accuracy such as bi--spectra and tri--spectra. So we provide the direct measurements of all perturbative terms using many realisations of simulations, which removes the uncertainty due to the unknown non--linear corrections. The systematic caused by the randomness of the peculiar velocity field can be controlled as well. The theoretical expression of the FoG term contains two distinct parts; the  ``one--point'' FoG term which is independent of the separation vector between two different points, and the ``correlated'' FoG term which is given by an infinite expansion of the velocity auto correlation fields. The ``correlated'' FoG term is coherently expanded in the same order as other higher order polynomials, and the one--point FoG term is provided in the closed formula with a single scale independent parameter of velocity dispersion. We will prove that, when the mapping formula is expanded up to second order polynomials of $k\mu$, the measured one--point FoG term at different scales is consistently aligned on the one single curve which is close to a Gaussian function. When the directly measured non--linear corrections and the pre--determined FoG functional form are exploited, the estimated 2-dimensional (hereafter 2D) redshift space power spectrum is well reproduced up to $k\lesssim0.2\hompc$ within the statistical error of the future survey.

The paper is organized as follows. In Sec.~\ref{sec:theory}, we introduce our theoretical model of redshift distortion. Sec.~\ref{sec:measurement} presents the simulation measurements of individual terms of the formula. In Sec.~\ref{sec:fog}, the residual FoG term is measured and the best fitting function is studied. Finally, we conclude and discuss in Sec.~\ref{sec:discussion}.

\section{RSD formula}
\label{sec:theory}

Both inhomogeneous density and velocity fields at large scales are small perturbations to the homogeneous background of the universe. If the first order approximation of fluctuations dominates, the observed redshift space power spectrum could be expanded in two--dimensional space spanned by radial and transverse directions. 

The density field seen at the redshift space is distorted by the peculiar motions of particles, halos or galaxies, 
\beq
\bfs=\bfr+\frac{\bfv \cdot \hat{z}}{aH},
\eeq
where $\bfr$ and $\bfs$ denote vector distances in real and redshift spaces respectively, and $\bfv$, $a$ and $H$ are the physical peculiar velocity, the scale factor and the Hubble parameter. We choose $\hat{z}$ direction as the line-of-sight direction. Following the derivation of \cite{Taruya10}, the observed 2D power spectrum in redshift space is given by \cite{Taruya10},
\begin{equation}
P^{\rm(S)}(k,\mu)=\int d^3\bfx\,e^{i\,\bfk\cdot\bfx}
\bigl\langle e^{j_1A_1}A_2A_3\bigr\rangle\,, 
\label{eq:Pkred_exact}
\end{equation}
in which we define
\begin{eqnarray}
&j_1= -i\,k\mu ,\nonumber\\
&A_1=u_z(\bfr)-u_z(\bfr'),\nonumber\\
&A_2=\delta(\bfr)+\,\nabla_zu_z(\bfr),\nonumber\\
&A_3=\delta(\bfr')+\,\nabla_zu_z(\bfr'),\nonumber
\end{eqnarray}
where $\bfx=\bfr-\bfr'$, $\bfu\equiv-\bfv/(aH)$, $u_z$ is the radial direction component of $\bfu$, and $\mu$ denotes the cosine of the angle between $\bfk$ and the line of sight. Eq.~(\ref{eq:Pkred_exact}) adopts the plane parallel approximation and single-streaming approximation, beside which it describes a rigorous mapping from real space density clustering to redshift space density clustering. 

The pairwise velocity field, $A_1$, which appears in the exponential function of the mapping in Eq.~(\ref{eq:Pkred_exact}), will cause an indefinite series of higher order polynomials, as shown later. We rewrite the ensemble average $\langle e^{j_1A_1}A_2A_3\rangle$ in terms of the connected cumulants~\cite{Scoccimarro04,Taruya10} as
\bea
&\langle e^{j_1A_1}A_2A_3\rangle=
\exp \left\{\langle e^{j_1A_1}\rangle_c\right\}
\nonumber\\
&\qquad\times
\left[\langle e^{j_1A_1}A_2A_3 \rangle_c+ 
\langle e^{j_1A_1}A_2\rangle_c \langle e^{j_1A_1}A_3 \rangle_c \right], \nonumber
\eea
then Eq. (\ref{eq:Pkred_exact}) reduces to 
\bea
&P^{\rm(S)}(k,\mu)=\int d^3\bfx \,\,e^{i\bfk\cdot\bfx}\,\,
\exp \left\{\langle e^{j_1A_1}\rangle_c\right\}
\nonumber\\
&\quad\quad
\times\left[\langle e^{j_1A_1}A_2A_3 \rangle_c+ 
\langle e^{j_1A_1}A_2\rangle_c \langle e^{j_1A_1}A_3 \rangle_c \right].
\label{eq:Pkred_exact2}
\eea
The ``perturbative terms" after mapping, which will be Taylor expanded later, are collected inside the bracket in the second line of the equation, which includes the squeezing Kaiser terms, and other higher order perturbative polynomials. These higher order polynomials have either amplifying or damping effects to $P^{\rm (S)}(k,\mu)$. The prefactor $\exp \left\{\langle e^{j_1A_1}\rangle_c\right\}$ mainly describes the uncertainty due to the randomness of the velocity field. It is the source of the FoG effect.

In the conventional assumption, the FoG prefactor $\exp \left\{\langle e^{j_1A_1}\rangle_c\right\}$ is treated as a term which could be factored out of the spatial integration (e.g. \cite{Taruya10}). That means that $\exp \left\{\langle e^{j_1A_1}\rangle_c\right\}$ is independent of the separation vector $\bfx$. Then the observed power spectrum could be approximately described as,
\beq
P^{\rm (S)}(k,\mu)=D^{\rm FoG}(k\mu\sigma_z)P_{\rm perturbed}(k,\mu) \nonumber,
\label{eq:phenom}
\eeq
where $D^{\rm FoG}(k\mu\sigma_z)$ is the FoG term, in which $\sigma_z^2\equiv\left\langle u_z^2\right\rangle_c$ denotes the line-of-sight velocity dispersion, and the $P_{\rm perturbed}$ represents the Fourier transformation of the bracket terms in the integral of Eq.~(\ref{eq:Pkred_exact2}). 

Rigorously, the above assumption of $\exp \left\{\langle e^{j_1A_1}\rangle_c\right\}$ is questionable, as that it can be mathematically divided into two parts \cite{Taruya10,Zhangrsd}, one ``one--point'' part $D^{\rm FoG}_{\rm 1pt}$ consisting of only one--point velocity cumulants, and the other ``correlated'' part $D^{\rm FoG}_{\rm corr}$ which includes auto velocity field correlations. After Taylor expanding $e^{j_1A_1}$, the decomposition is shown as,
\begin{widetext}
\bea
\label{eq:fullfog1}\exp\left\{\langle e^{j_1A_1}\rangle_c\right\}&=&\exp\left\{\sum_{n=1}^{\infty}j_1^n\frac{\langle A_1^n\rangle_c}{n!}\right\}
=\exp\left\{\sum_{n=1}^{\infty}j_1^{2n}\frac{\langle (u_z(\bfr)-u_z(\bfr'))^{2n}\rangle_c}{(2n)!}\right\}  \\
&=&\label{eq:fullfog2}\exp\left\{\sum_{n=1}^{\infty}j_1^{2n}\frac{2\langle u_z(\bfr)^{2n}\rangle_c}{(2n)!}\right\} 
\times \exp\left\{\sum_{n=1}^{\infty}j_1^{2n}\frac{\langle (u_z(\bfr)-u_z(\bfr'))^{2n}\rangle_c -\langle u_z(\bfr)^{2n}\rangle_c-\langle u_z(\bfr')^{2n}\rangle_c}{(2n)!}\right\} \\
&=&\label{eq:fullfog3}D^{\rm FoG}_{\rm 1pt}(k\mu)\times D^{\rm FoG}_{\rm corr}(k\mu,\bfx).
\eea
The odd power terms in Eq.~(\ref{eq:fullfog2}) are nullified due to the symmetric characteristic of the velocity field, $\left\langle u_z(\bfr)^\alpha u_z(\bfr')^\beta \right\rangle_c=0$ when $\alpha+\beta=2n+1$. Two exponential terms in Eq.~(\ref{eq:fullfog2}) are defined in Eq.~(\ref{eq:fullfog3}) in order by $D^{\rm FoG}_{\rm 1pt}$ and $D^{\rm FoG}_{\rm corr}$, which could be reformed as,
\bea
\label{eq:fog_non_local}
D^{\rm FoG}_{\rm 1pt}(k\mu)&\equiv & \exp\left\{\sum_{n=1}^{\infty}j_1^{2n}\frac{2\langle u_z(\bfr)^{2n}\rangle_c}{(2n)!}\right\} =\exp\left\{j_1^2\sigma_z^2+\sum_{n=2}^{\infty}j_1^{2n}\frac{2\langle u_z(\bfr)^{2n}\rangle_c}{(2n)!}\right\},\\
\label{eq:fog_local}
D^{\rm FoG}_{\rm corr}(k\mu,\bfx)&\equiv & \exp\left\{\sum_{n=1}^{\infty}j_1^{2n}\frac{\langle (u_z(\bfr)-u_z(\bfr'))^{2n}\rangle_c -\langle u_z(\bfr)^{2n}\rangle_c-\langle u_z(\bfr')^{2n}\rangle_c}{(2n)!}\right\} \nonumber \\
&=&\exp\left\{-j_1^2\langle u_z(\bfr)u_z(\bfr')\rangle_c +\sum_{n=2}^{\infty}j_1^{2n}\frac{\langle (u_z(\bfr)-u_z(\bfr'))^{2n}\rangle_c -\langle u_z(\bfr)^{2n}\rangle_c-\langle u_z(\bfr')^{2n}\rangle_c}{(2n)!}\right\}.
\eea
\end{widetext}

This classification of FoG assists us to isolate the FoG term originated from one--point velocity distribution function. The FoG part originated from the spatial correlation $\bfx$ remains in the integration with other perturbative terms. It will be expanded coherently in the same order as higher order polynomials of density--velocity cross correlations. And the $D^{\rm FoG}_{\rm 1pt}$ is safely factored out of the integration, and as shown later, it could be parameterized with a single parameter of $\sigma_z$, which by definition is a constant. Then the conventional expression of the observed spectrum in Eq.~(\ref{eq:phenom}) is rewritten as,
\bea
&P^{\rm(S)}(k,\mu)=D^{\rm FoG}_{\rm 1pt}(k\mu)\int d^3\bfx \,\,e^{i\bfk\cdot\bfx}D^{\rm FoG}_{\rm corr}(k\mu,\bfx)
\nonumber\\
&\quad\quad
\times  \left[\langle e^{j_1A_1}A_2A_3 \rangle_c+ 
\langle e^{j_1A_1}A_2\rangle_c \langle e^{j_1A_1}A_3 \rangle_c \right].
\label{eq:Pkred_exact3}
\eea

Defining the reduced cumulants of the velocity field, $K_i\equiv \left\langle u^i\right\rangle_c/\sigma_z^i$, the one--point FoG term $D^{\rm FoG}_{\rm 1pt}$ could be written as \cite{Zhangrsd},
\beq
\label{eq:fog_non_local2}
D^{\rm FoG}_{\rm 1pt}(k\mu)=\exp\left\{j_1^2\sigma_z^2+2\sum_{n=2}^{\infty}j_1^{2n}\sigma_z^{2n}\frac{K_{2n}}{(2n)!}\right\},
\eeq
where the higher order reduced velocity cumulants $K_{2n}$ ($n \geq 2$) indicate the non-Gaussianity of the velocity field, which is generated by non-linear structure formation. In \cite{Zheng13}, authors measured $\sigma_z^2$ and higher order $K_{2n}$ from simulations, together with the exact one--point FoG term by integrating the measured velocity Probability Distribution Function. 
Here, $\sigma_z^2$ is not only the dispersion of small scale random velocity, but the dispersion of the whole velocity field, which could be expressed as the integral of the velocity power spectrum,
\beq
\label{eq:sigmaz}
\sigma_z^2=\frac{1}{3}\int \frac{dk}{2\pi^2}\Ptt\,.
\eeq
However, in this manuscript, $\sigma_z^2$ is assumed to be theoretically unknown, and parameterised to be fitted with all other cosmological parameters.
The other FoG term and other perturbative parts in Eq.~(\ref{eq:Pkred_exact3}) are indefinite to be expressed by order expansion. In the regime of small $j_1$, $e^{j_1A_1}$ is expanded in terms of $j_1$ as,
\bea
&& D^{\rm FoG}_{\rm corr}(k\mu,\bfx)\left[\langle e^{j_1A_1}A_2A_3 \rangle_c+\langle e^{j_1A_1}A_2\rangle_c \langle e^{j_1A_1}A_3 \rangle_c \right] \nonumber \\
&&\simeq j_1^0\langle A_2A_3\rangle_c + j_1^1\langle A_1A_2A_3\rangle_c \nonumber\\
&&+j_1^2\Bigl\{\langle A_1A_2\rangle_c\langle A_1A_3\rangle_c + \frac{1}{2}\,\langle A_1^2A_2A_3\rangle_c-\langle u_z u_z'\rangle_c\langle 
A_2A_3\rangle_c \Bigr\} \nonumber\\
&&+\mathcal{O}(j_1^3)\,,
\label{eq:expansion}
\eea
where we truncate the expansion at $j_1^2$ order. Then the zeroth order term, $\langle A_2A_3\rangle_c$, representing the squeezing effect, corresponds to $P_{\delta\delta}+2\mu^2P_{\delta\Theta}+\mu^4P_{\Theta\Theta}$ \cite{Scoccimarro04} with the velocity divergence $\Theta\equiv -\nabla\cdot \bfv/(aH)=\nabla \cdot \bfu$. The higher order terms are defined as,
\begin{eqnarray}
  A(k,\mu)&=& j_1\,\int d^3\bfx \,\,e^{i\bfk\cdot\bfx}\,\,\langle A_1A_2A_3\rangle_c,\nonumber\\
  B(k,\mu)&=& j_1^2\,\int d^3\bfx \,\,e^{i\bfk\cdot\bfx}\,\,\langle A_1A_2\rangle_c\,\langle A_1A_3\rangle_c,\nonumber\\
  T(k,\mu)&=& \frac{1}{2} j_1^2\,\int d^3\bfx \,\,e^{i\bfk\cdot\bfx}\,\,\langle A_1^2A_2A_3\rangle_c,\nonumber \\
  F(k,\mu)&=& -j_1^2\,\int d^3\bfx \,\,e^{i\bfk\cdot\bfx}\,\,\langle u_z u_z'\rangle_c\langle A_2A_3\rangle_c, \nonumber
\end{eqnarray}
Here $A$, $B$ and $T$ terms are originated from higher order density--velocity cross correlation polynomials, and $F$ terms comes from higher order velocity auto correlations.

Finally we derive the RSD model in coherent order expansion of $k\mu$ \cite{Taruya10},
\bea
\label{eq:Pkred_final}
P^{\rm (S)}(k,\mu)&=&D^{\rm FoG}(k\mu\sigma_z)P_{\rm perturbed}(k,\mu) \nonumber\\ 
&=&D^{\rm FoG}(k\mu\sigma_z)[P_{\delta\delta}+2\mu^2P_{\delta\Theta}+\mu^4P_{\Theta\Theta}  \\
&&+A(k,\mu)+B(k,\mu)+T(k,\mu)+F(k,\mu)] \nonumber.
\eea
This model is first derived in TNS model paper \cite{Taruya10}, in which they ignored T and F terms. So one goal of this paper is to investigate the improvement of including T and F terms in the model.

From now on, we drop the subscript of ``1pt", and use $D^{\rm FoG}(k\mu\sigma_z)$ to represent the one--point FoG effect. Although $D^{\rm FoG}(k\mu\sigma_z)$ function is expected to be Gaussian \cite{Zheng13}, the exact functional form is assumed to be unknown in this paper. The higher order truncation effect is absorbed into $D^{\rm FoG}(k\mu\sigma_z)$. Consequently $\sigma_z$ represents effects from the summation of velocity cumulants and the truncated higher order polynomials, remaining unknown but a constant. We test FoG function in the order expansion in terms of $(k\mu\sigma_z)^2$, 
\begin{eqnarray}
D^{\rm FoG}(k\mu\sigma_z)=
e^{-k^2\mu^2\sigma_z^2/2+\sum_{n=1} E_{2n} (k\mu\sigma_z)^{2n}} 
\label{eq:fog_phenom}
\end{eqnarray}
where $\sigma_z$ is set to be a free parameter fitted by simulation measurements, and $E_{2n}$ represents the free coefficient of the higher order contribution which will be marginalised over. In this manuscript, the perturbative terms are expanded up to $j_1^2$ order, and the only undetermined parameter becomes $\sigma_z$ with the given FoG functional form. If this order truncation approximation is valid, $\sigma_z$ remains a constant. When it becomes scale dependent, the RSD model is considered to be contaminated by higher order $j_1^n(n \geq 3)$ terms.

\section{measurements of the perturbative terms}
\label{sec:measurement}

In this section, we provide the individual measurements for all perturbative terms in Eq.~(\ref{eq:Pkred_final}), using the N-body simulations. In Sec.~\ref{sec:fog}, the summation of them will be divided from the measured 2D redshift space power spectrum to obtain the one--point FoG term measurements. This helps us remove the uncertainty due to the non--linear evolution which can not be perfectly predicted by perturbation theory. Though the non--linear spectra at zeroth order of $j_1$ are well known both by analytic formulation and by simulation, higher order polynomials have been calculated only theoretically, and have not yet been confirmed directly for each term from simulations.

Our work is based on 100 realisations of N--body dark matter simulations (Minji Oh et.al 2016, in preparation). Each simulation is run by the cosmological simulation code GADGET2 \cite{Springel05}, with box-size $L_{\rm box}=1.89\,h^{-1}$Gpc and $N_{\rm p}=1024^3$ particles. The box-size is chosen to mimic a similar survey volume as DESI will observe between $z=0.8$ and $z=1.0$ \cite{DESIwhite}. All realisations are generated with the same cosmological model, so 100 simulations could help us identify the systematic contamination at  10$\%$ level of the targeted signal to noise for DESI survey. We assume a LCDM cosmology with Gaussian initial condition and flat space. The cosmological parameters are identical to PLANK15 results \cite{PLANK2015}. The initial conditions are generated by the 2LPT code \cite{2LPT} at $z=49$. We mainly analyse 4 snapshots in this paper, namely $z=0.5$, 0.9, 1.5, and 3.0. The detailed simulation parameters are listed in Table \ref{tab:simulation}.

\begin{table}[t]
\scriptsize
\begin{tabular}{@{}lll}
\hline\hline
parameter & physical meaning & value \\
\hline
$\Omega_m$  & present fractional matter density & $0.3132$ \\
$\Omega_{\Lambda}$ & $1-\Omega_m$ & $0.6868$ \\
$\Omega_b$ & present fractional baryon density & $0.049$\\
$h$ & $H_0/(100$~km~s$^{-1}$Mpc$^{-1})$ & $0.6731$ \\
$n_s$ & primordial power spectral index & $0.9655$ \\
$\sigma_{8}$ & r.m.s. linear density fluctuation & $0.829$ \\
\hline
$L_{\rm box}$ & simulation box size & 1890~$h^{-1}$Mpc\\
$N_{\rm p}$ & simulation particle number & $1024^3$\\
$m_{\rm p}$ & simulation particle mass & $5.46\times 10^{11}h^{-1}M_{\odot}$\\
\hline
$N_{\rm snap}$ & number of output snapshots & $13$ \\
$z_{\rm ini}$ & redshift when simulation starts & $49.0$ \\
$z_{\rm final}$ & redshift when simulation finishes & $0.0$ \\
\hline

\end{tabular}
\caption{The parameters and technical specifications of the N-body simulations for this work.}
\label{tab:simulation}
\end{table}

\bfi{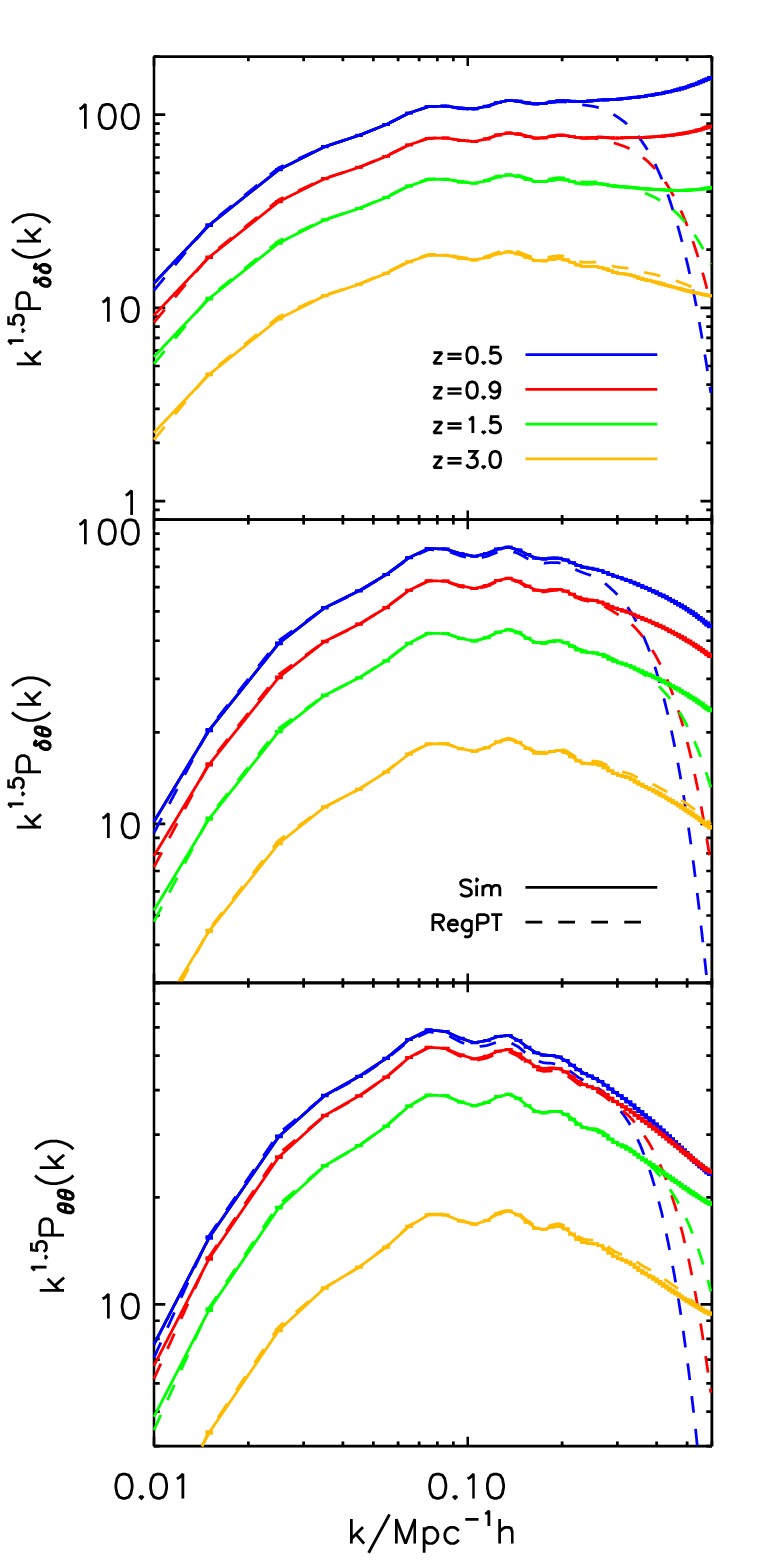}
\caption{We present the measured $\Pdd$, $\Pdt$, and $\Ptt$ versus the theoretical $\Pdd$, $\Pdt$, and $\Ptt$ at $z=0.5$, 0.9, 1.5, and 3.0., as solid and dash curves respectively. The top, middle and bottom panels represent tests on $\Pdd$, $\Pdt$, and $\Ptt$ respectively. The measured $\Pdd$, $\Pdt$, and $\Ptt$ are mean values out of 100 simulations, and the small error bars on the solid curves denote the standard errors of the mean, $\sigma_{\rm mean}=\sigma/\sqrt{N}$, with $\sigma$ being the sample standard deviation, $N=100$ being the sample number. The dashed lines show the predictions from RegPT \cite{RegPT}.}
\label{fig:pk}
\efi

In the following subsections, we describe the methodology to directly measure each term in Eq.~(\ref{eq:Pkred_final}), which includes three $j_1^0$ terms, one $j_1^1$ term and three $j_1^2$ terms. We separately compute the averages of all those terms from 100 realisations, which are assumed to be sufficiently close to the true values. The averages from the limited number of realisations leaves the small deviations, but the difference is as small as 10\% of the statistical uncertainty of the given simulation volume. Although some terms could be expressed as the integrations of the higher order correlation functions larger than two points, as shown in Sec.~\ref{sec:theory}, all perturbative terms  are essentially defined at only two distinct points of $\bfr$ and $\bfr'$, whose measurements turn out to be calculations of only two point correlation functions.  Mathematically, all terms can be decomposed into three basic elements of $\delta(\bfr)$, $u_z(\bfr)$ and $\nabla_z u_z(\bfr)$. So first we use the Nearest Grid Point (NGP) method to sample the density field on regular $512^3$ girds and use the Nearest Particle (NP) \cite{Zheng13} method to sample the velocity fields on regular $512^3$ grids. Then the combination of three elements at each point is computed, finally we construct two specific fields by summing the basic elements and make the two point correlation functions of these two fields to complete the measurements.   

Though we correct the window function effect in measured real space and redshift density power spectra and real space density-velocity cross power spectrum, all measurements still contain systematic errors due to finite particle number and grid size \cite{Jing05,Zhang14,Zheng14a}. But since our simulations have high enough particle number density and small enough grid size, all these numerical effects are negligible at $1\%$ level for the scientific goal of this work, at least within our interested scales, e.g., $k<0.2h/$Mpc
\footnote{Main uncertainty comes from the volume-weighted velocity field estimation. We make convergence tests by comparing the velocity power spectrum measured from a fraction of randomly selected particles with that of all the particles. This gives good estimation of the systematic error of our sampled velocity field. For more details, please refer to \cite{Zheng14a} and the Appendix of~\cite{Zheng13}.}. We also compare the simulation calculations with perturbative predictions to show the break-down scales of the perturbation theories in this section.

\subsection{$j_1^0$ order: power spectra}
\label{subsubsec:j0}

In this subsection, we calculate the zero's order terms in the $j_1$ expansion, using both perturbation theory and direct measurement from the simulations. There is no other higher order polynomials but the simple Kaiser terms of $\Pdd$, $\Pdt$, and $\Ptt$ in this lowest order.

First, these Kaiser terms are calculated using perturbation theory. The odd UV behaviour appears in the standard perturbation theory caused by the incomplete expansion. It is improved by incorporating an appropriate UV regularization, which removes all bad UV sectors, dubbed as the regularized perturbation theory \cite{RegPT} (hereafter {\tt RegPT}). The {\tt RegPT} theory is based upon a multipoint propagator expansion which applies to both density and velocity field multiple propagators to encapsulate the non-perturbative properties of gravitational effect. We apply the {\tt RegPT} for deriving theoretical prediction of $j_1^0$ order terms, and present the results as dash curves in Fig.~\ref{fig:pk}. The predictions of $\Pdd$, $\Pdt$, and $\Ptt$ using {\tt RegPT} are shown in the top, middle and bottom panels respectively at four different redshifts of $z=0.5$, 0.9, 1.5, and 3.0.

Next, the $\Pdd$, $\Pdt$, and $\Ptt$ spectra are directly measured using simulations, which are represented by solid curves in the top, middle and bottom of Fig.~\ref{fig:pk} respectively. Theoretical prediction fails at the regime in which UV behaviour becomes dominant. The difference between solid and dash curves is explained by this reason. The theoretical prediction for $\Pdd$ agrees with the measurement at $k\la 0.2\hompc$ at low redshifts, but the $\Ptt$ prediction breaks down at $k\ga 0.1\hompc$, which causes the inaccurate prediction of the observed spectrum at $\mu\rightarrow 1$ limit. If we are interested in RSD model at $k\ga 0.1\hompc$, more precise theoretical prediction is demanded. In this manuscript, the measured $j_1^0$ order terms from simulations are used, and our results are free from UV issues.

\subsection{$j_1^1$ order: $A$ term}
\label{subsubsec:j1}

\begin{figure}
\centering
\includegraphics[width=1.0\columnwidth]{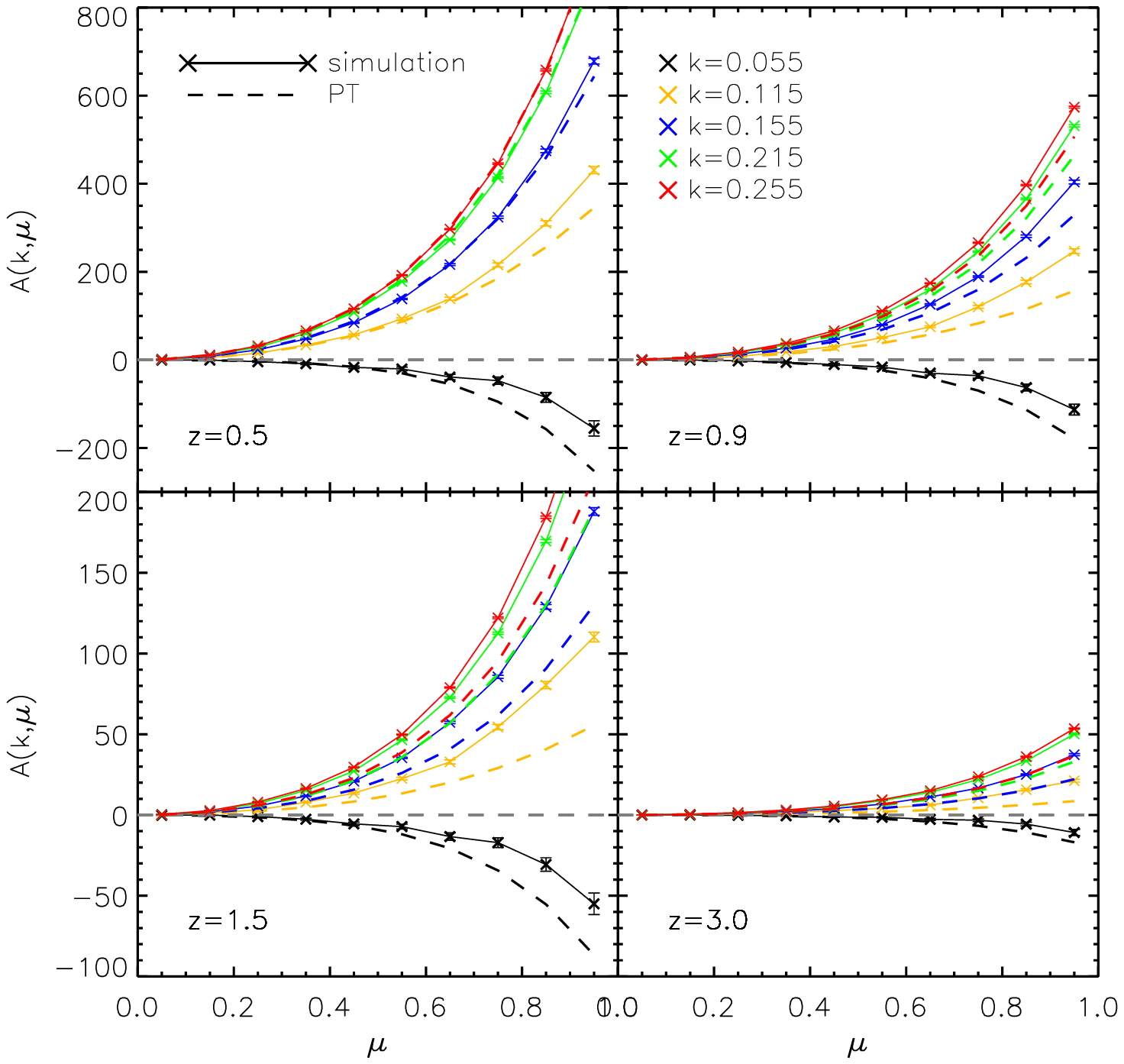}
\caption{$A(k,\mu)$ term measured from 100 N-body simulations at $z=0.5$, 0.9, 1.5, and 3.0. Different colors represent different $k$ bins. The solid lines are the mean value averaged over 100 simulations, and the error bars are the standard errors of the mean, $\sigma_{\rm mean}=\sigma/\sqrt{N}$, with $\sigma$ being the sample standard deviation, $N=100$ being the sample number. The dashed lines show the predictions from standard perturbation theory \cite{Taruya10}. We caution that the y-axis ranges of up and bottom panels are different. }
\label{fig:Aterm}
\end{figure}

The $A(k,\mu)$ in Eq.~(\ref{eq:Pkred_final}) is a leading $j_1^1$ order in polynomial expansion generated by non--linear mapping of density--velocity cross--correlation. There is no corresponding $j_1^1$ order term in the expansion of velocity auto--correlation mapping. The $A(k, \mu)$ term is described as \cite{Taruya10},
\bea
\label{eq:Aterm}
A(k,\mu)&=& j_1\,\int d^3\bfx \,\,e^{i\bfk\cdot\bfx}\,\,\langle A_1A_2A_3\rangle_c\nonumber\\
&=&j_1\,\int d^3\bfx \,\,e^{i\bfk\cdot\bfx}\,\,\langle (u_z-u_z')\\ \nonumber
&&\times(\delta+\nabla_zu_z)(\delta'+\nabla_zu_z')\rangle_c\,\\
\label{eq:Aterm_bi}
&=&(k\mu)\int \frac{d^3\bf{p}}{(2\pi)^3} \frac{p_z}{p^2}\lbrace B_\sigma(\bf{p},\bfk-\bf{p},-\bfk) \nonumber \\
&&-B_\sigma(\bf{p},\bfk,-\bfk-\bf{p})\rbrace,
\eea
where the bispectrum $B_\sigma$ is defined by
\bea
&&\left\langle \theta(\bfk_1)
\left\{\delta(\bfk_2)+\,\frac{k_{2z}^2}{k_2^2}\theta(\bfk_2)\right\}
\left\{\delta(\bfk_3)+\,\frac{k_{3z}^2}{k_3^2}\theta(\bfk_3)\right\}
\right\rangle
\nonumber\\
&&=(2\pi)^3\delta_D(\bfk_1+\bfk_2+\bfk_3)\,B_\sigma(\bfk_1,\bfk_2,\bfk_3).
\label{eq:cross_bi}
\eea
The theoretical solution of $A(k,\mu)$ is calculated consistently with theoretical $j_1^0$ terms derived using {\tt RegPT} scheme which are presented as dash curves in Fig.~\ref{fig:pk}. The level in the $A(k,\mu)$ perturbation is corresponding to the tree level, in which $j_1^0$ order terms are selected to compute this level by incorporating one loop level. Then the perturbative expansion in $B_\sigma$ is truncated at the leading order, and other higher order levels are ignored (see the Appendix of \cite{Taruya10} for details). The theoretical solution is presented by dash curves in four panels in Fig.~\ref{fig:Aterm}. Each panel represents the results at different redshifts $z=0.5$, 0.9, 1.5 and 3.0.

\begin{figure}
\centering
\includegraphics[width=1.0\columnwidth]{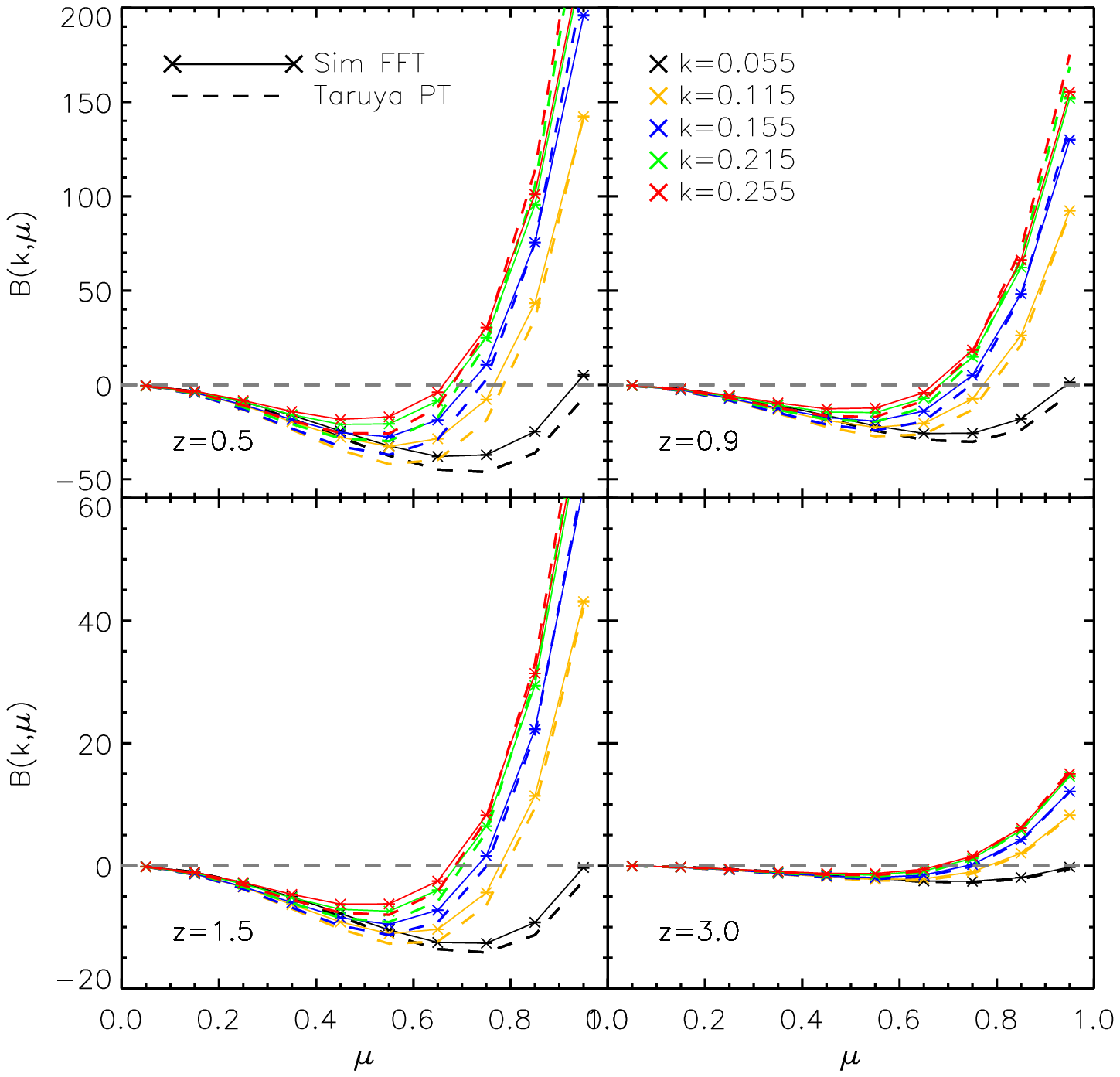}
\caption{Similar with Fig. \ref{fig:Aterm}, but for $B(k,\mu)$ term. We caution that the y-axis ranges of up and bottom panels are different.}
\label{fig:Bterm}
\end{figure}

While the theoretical solution of $A(k,\mu)$ is derived by integrating bi--spectra in Fourier space using Eq.~(\ref{eq:Aterm_bi}), it is a time--consuming procedure when $A(k,\mu)$ is computed numerically from simulations. Instead, we exploit the expression given in Eq.~(\ref{eq:Aterm}), which is effectively decomposed into two point functions in the configuration space. All perturbative fields of $\delta(\bfr)$, $u_z(\bfr)$, $\nabla_z u_z(\bfr)$, $\delta u_z(\bfr)$, and $u_z\nabla_z u_z(\bfr)$ are separately measured to be combined at two different points. The combined fields at both $\bfr$ and $\bfr'$ are cross--correlated appropriately, and the measured pairs in the configuration space are transformed into the Fourier space. We collect all Fourier components to provide the numeric $A(k,\mu)$, which is presented as solid curves in Fig.~\ref{fig:Aterm} at diverse redshifts.

We compare the theoretical and numerical results in Fig.~\ref{fig:Aterm} using characteristic scales from linear to non--linear regimes of $k=(0.055,0.115,0.155,0.215,0.255)\hompc$. Both agree at small $\mu\rightarrow 0$ limit, but deviates to each other at $\mu\rightarrow 1$. It is difficult to explain the origin of the differences, as higher order perturbative theory is not well understood beyond two point functions. However, the theoretical prediction even at the tree level is quite consistent with the measurement, which explains most successful applications of the improved theoretical model for observations at low redshifts. Note that the effect of $j_1^1$ order contributes  differently to the observed power spectrum, in that the enhancement at small $k$ modes and the decrement at high $k$ modes, which are numerically confirmed here. In this work, because the numerical $j_1^0$ order solutions are adopted, the numeric $A(k,\mu)$ is counted.

\subsection{$j_1^2$ order: $B$, $T$ and $F$ terms}
\label{subsubsec:j2}

Three terms of $B(k,\mu)$, $T(k,\mu)$ and $F(k,\mu)$ in Eq.~(\ref{eq:Pkred_final}) represent the complete components in the expansion of $j_1^2$ order. The $B(k,\mu)$ and $T(k,\mu)$ are provided by non--linear mapping of density--velocity cross--correlations, and the $F(k,\mu)$ comes from the mapping of velocity auto--correlation $\langle u_z u_z'\rangle_c$ pairs which are generated from the FoG effect. In this subsection, we calculate these terms numerically, and compare them with theoretical predictions if available.

\begin{figure}
\centering
\includegraphics[width=1.0\columnwidth]{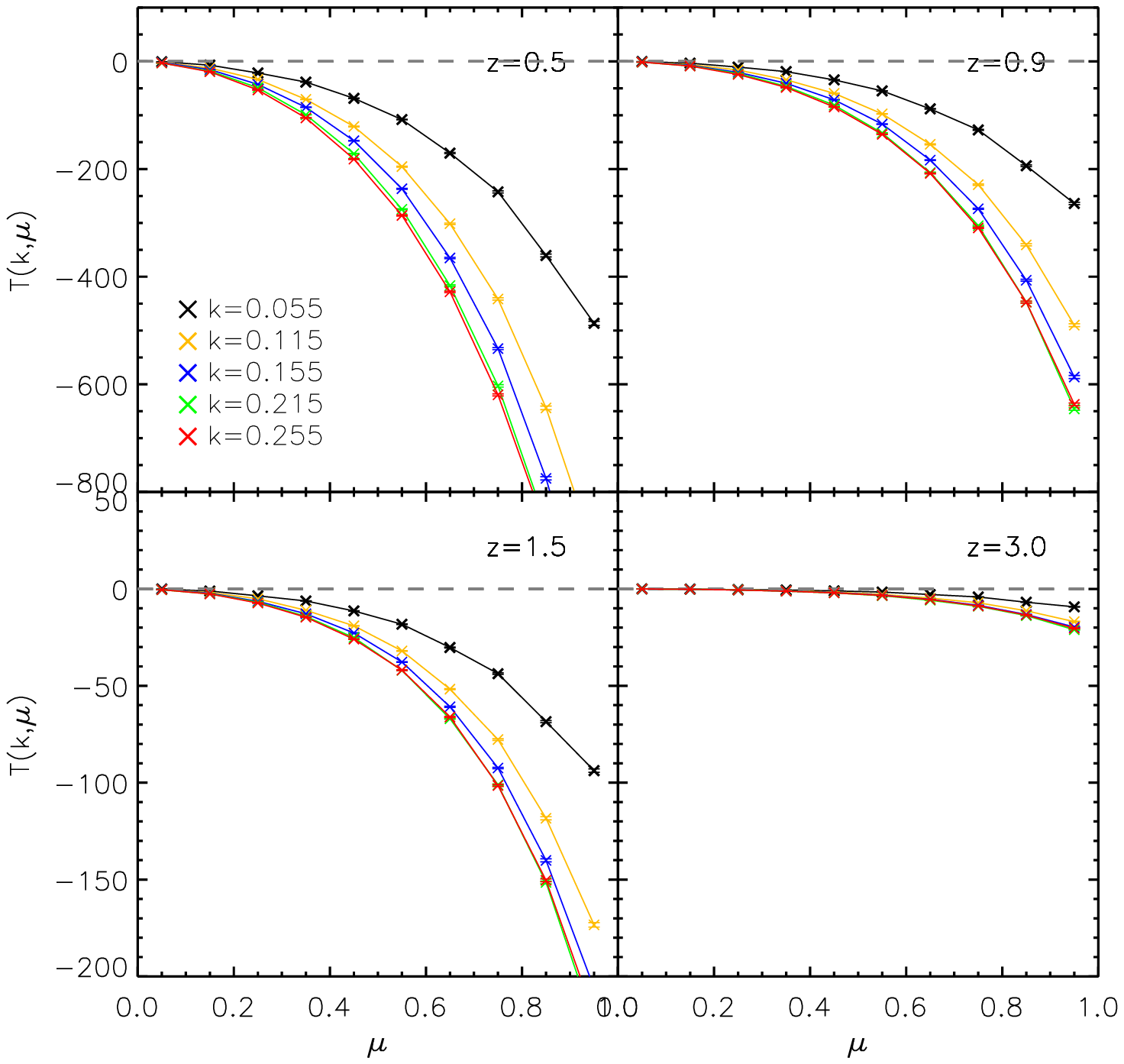}
\caption{Similar with Fig. \ref{fig:Aterm}, but for $T(k,\mu)$ term. We caution that the y-axis ranges of up and bottom panels are different.}
\label{fig:Dterm}
\end{figure}

The expression of $B(k,\mu)$ is given by \cite{Taruya10},
\bea
\label{eq:Bterm}
B(k,\mu)&=& j_1^2\,\int d^3\bfx \,\,e^{i\bfk\cdot\bfx}\,\,\langle A_1A_2\rangle_c\,\langle A_1A_3\rangle_c\nonumber\\
&=&j_1^2\,\int d^3\bfx \,\,e^{i\bfk\cdot\bfx}\,\,\langle (u_z-u_z')(\delta+\nabla_zu_z)\rangle_c\,\nonumber\\
&&\times \langle (u_z-u_z')(\delta'+\nabla_zu_z')\rangle_c.
\eea
Then the numeric solution for $B(k,\mu)$ is straightforward. Again, we compute both two point correlation functions of $\langle A_1A_2\rangle_c$ and $\langle A_1A_3\rangle_c$ to be combined into the integrand in terms of $\bfx$. The Fourier transformation of that integrand provides us the correct solution of $B(k,\mu)$. Unlike the case of $A(k,\mu)$, the alternative expression of $B(k,\mu)$ is not complicated to be used for numeric calculation. The $B(k,\mu)$ could also be expressed in the Fourier space as,
\bea
\label{eq:Bterm2}
B(k,\mu)= (k\mu\,f)^2\int \frac{d^3\bfp}{(2\pi)^3} F(\bfp)F(\bfk-\bfp), \\
F(\bfp)=\frac{p_z}{p^2}\left\{ \Pdt(p)+f\,\frac{p_z^2}{p^2}\,\Ptt(p)\,\right\}. \nonumber
\eea
If all power spectra of density and velocity fields are directly measured as in Sec.~\ref{subsubsec:j0}, the integrand in Fourier space can be derived correctly. 
We compare both measured results from Eq.~(\ref{eq:Bterm}) and Eq.~(\ref{eq:Bterm2}), and they are consistent with each other.  This indirectly confirms that our method applied for calculating $B(k,\mu)$ in the configuration space is correct. The measured $B(k,\mu)$ are presented as solid curves in Fig.~\ref{fig:Bterm}. We see that the amplitude of $B(k,\mu)$ term is $3\sim4$ times smaller than that of $A(k,\mu)$ term. 

\begin{figure}
\centering
\includegraphics[width=1.0\columnwidth]{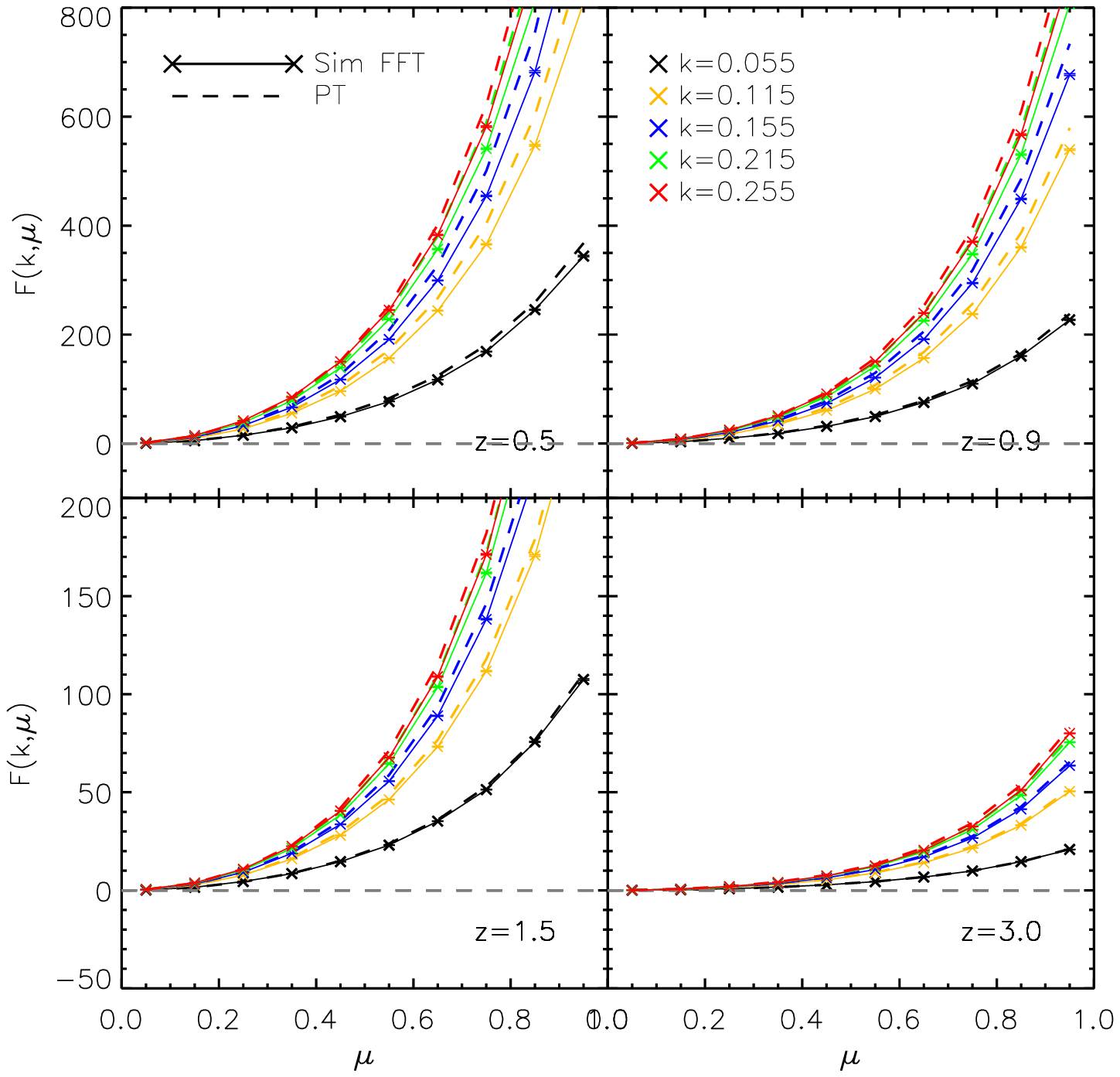}
\caption{Similar with Fig. \ref{fig:Aterm}, but for $F$ term. We caution that the y-axis ranges of up and bottom panels are different.}
\label{fig:Cterm}
\end{figure}

\bfig{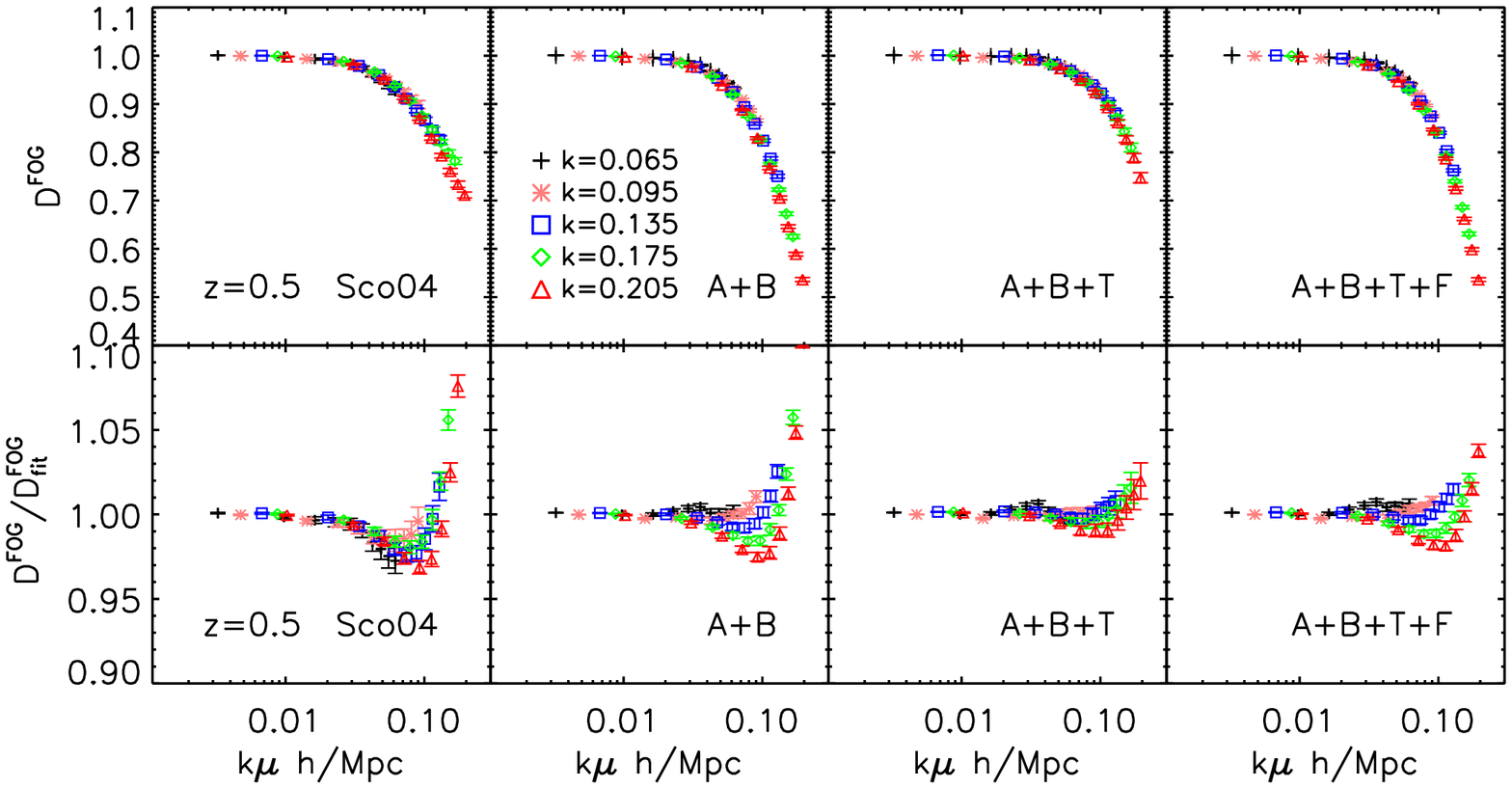}
\caption{{\it (Top panel)} The measured residual FoG's are presented at $z=0.5$, which is explained in the section~\ref{subsec:bestfog}. The selected scales are $k=(0.065,0.095,0.135,0.175,0.205)\mpcoh$. The residual FoG is computed by taking ratio described in Eq.~\ref{eq:fogfig}. The $P^{(S)}_{\rm perturbed}(k)$ include; 1) Kaiser terms only, 2) Kaiser terms and $A+B$, 3) Kaiser terms and $A+B+T$, and 4) Kaiser terms and $A+B+T+F$, from the first to the fourth columns. {\it (Bottom panel)} The best fit residual FoG $D^{\rm FOG}$ is derived using the correspondent model,  with free $\sigma_z$ and $E_{2n}=0$. We present the fractional ratio between the measured $D^{\rm FOG}$ and $D^{\rm FOG}_{\rm fit}$.} 
\label{fig:FoGres}
\efig

The theoretical prediction of $B(k,\mu)$ is derived using tree level perturbative theory, by substituting the linear density power spectrum into Eq.~(\ref{eq:Bterm2}), which is presented as dash curves in Fig.~\ref{fig:Bterm}. The agreement between the theoretical and numerical $B(k,\mu)$ is impressive, although there is slight difference observed at low redshifts. 
It is predicted that it decays the observed spectrum at small $\mu$ and enhances it at large $\mu$, which is confirmed numerically as well.

The expression of $T(k,\mu)$ in Eq.~\ref{eq:Pkred_final} is given by \cite{Taruya10},
\bea
T(k,\mu)&=& \frac{1}{2} j_1^2\,\int d^3\bfx \,\,e^{i\bfk\cdot\bfx}\,\,\langle A_1^2A_2A_3\rangle_c \nonumber \\
&=& \frac{1}{2} j_1^2\,\int d^3\bfx \,\,e^{i\bfk\cdot\bfx}\,\,\left\{\langle A_1^2A_2A_3\rangle\right. \nonumber\\
&&\left.-\langle A_1^2\rangle\langle A_2A_3\rangle-2\langle A_1A_2\rangle\langle A_1A_3\rangle\right\} \nonumber \\
&=&\frac{1}{2} j_1^2\,\int d^3\bfx \,\,e^{i\bfk\cdot\bfx}\,\,\left\{\langle (u_z-u_z')^2 \right.\nonumber \\
&&\times(\delta+\nabla_zu_z)(\delta'+\nabla_zu_z')\rangle\nonumber\\
&&\left.-\langle A_1^2\rangle\langle A_2A_3\rangle-2\langle A_1A_2\rangle\langle A_1A_3\rangle\right\}.
\eea
We apply the same numerical method as $A(k,\mu)$ calculation to derive $T(k,\mu)$ here. We construct the combined fields at $\bfr$ and $\bfr'$ separately, and compute the correlation functions in the configuration space. Those terms are collected to be transformed into the Fourier space, which are presented as solid curves in Fig. \ref{fig:Dterm}. The detailed test is presented in Appendix~\ref{app}.

The numerical results are shown in Fig.~\ref{fig:Dterm}, showing that the $T(k,\mu)$ contributes to the suppression of the observed spectrum at all scales and pair orientations. It is remarkable that this contribution is comparable to terms of $A(k,\mu)$ and $B(k,\mu)$ which are key elements in the improved theoretical RSD model \cite{Taruya10}. This numerical results are different from the perturbative calculations in \cite{Taruya13}, which shows that $T$ term is small compared to $A$ or $B$ term. And we regard to difference to the non--linear effect which may not be included in the perturbative calcualtions. Since $T$ terms effectively suppresses the observed spectrum in redshift space, it can be considered to be a part of FoG suppression. However, FoG suppression to the observed spectrum is coherent in terms of scales, which is a distinct feature from that of $T(k,\mu)$ suppression. If $T(k,\mu)$ is absorbed into the FoG effect, then the residual FoG will not present the scale--independence in terms of $k$.

The origin of $F(k,\mu)$ in Eq.~(\ref{eq:Pkred_final}) is different from other $j_1^2$ terms which are only parts of perturbative higher order Kaiser terms. The local FoG effect contains the higher order velocity correlations depending on the separation of two points. This velocity auto--correlation polynomials are indefinite as well, which needs to be expanded. It is also chosen to be expanded it in terms of $j_1$, then both the perturbative terms and local FoG polynomials are expanded coherently. The leading order of this effect appears as the multiplication between $\langle u_zu_z' \rangle_c$ and $\langle A_2A_3\rangle_c$ in $j_1^2$, which is given by \cite{Taruya10},
\bea
F(k,\mu)&=& -j_1^2\,\int d^3\bfx \,\,e^{i\bfk\cdot\bfx}\,\,\langle u_z u_z'\rangle_c\langle A_2A_3\rangle_c \nonumber\\
&=&-j_1^2\,\int d^3\bfx \,\,e^{i\bfk\cdot\bfx}\,\,\langle u_z u_z'\rangle_c  \nonumber \\
&&\times\langle (\delta+\nabla_zu_z)(\delta'+\nabla_zu_z')\rangle_c.
\label{eq:Cterm_FFT}
\eea
This equation is numerically computed from simulations using the similar methods described as $B(k,\mu)$ term calculation. Again, the alternative tree level expression in Fourier space is given by \cite{Taruya10},	
\bea
\label{eq:Cterm_integral}
F(k,\mu)&=&(k\mu\,f)^2\int\frac{d^3\bfp d^3\bfq}{(2\pi)^3}\,\delta_{\rm D}(\bfk-\bfp-\bfq)\,\frac{\mu_p^2}{p^2}\Ptt(p) \nonumber\\
&\times & \left\{\Pdd(q)+2\,f\,\mu_q^2\,\Pdt(q)+f^2\,\mu_q^4\,\Ptt(q)\right\} \\
&\simeq&  (k\mu\,f)^2\int\frac{d^3\bfp d^3\bfq}{(2\pi)^3}\,\delta_{\rm D}(\bfk-\bfp-\bfq)\nonumber\\
&&\times\frac{\mu_p^2}{p^2}\left(1+f\,\mu_q^2\right)^2P_{\rm lin}(p)P_{\rm lin}(q).
\label{eq:Cterm_PT}
\eea
We cross--check that the results from Eq.~(\ref{eq:Cterm_FFT}) and ~(\ref{eq:Cterm_integral}) are consistent with each other, which confirms our numerical calculation.

The structure of $F(k,\mu)$ is similar to $B(k,\mu)$ case in which the simple tree level calculation agrees with the direct measurement using simulations. The directly measured $F(k,\mu)$ is presented as solid curves in Fig.~\ref{fig:Cterm}, and the prediction using Eq.~(\ref{eq:Cterm_PT}) is presented as dash curves. Both are consistent. Note that the effect of $F(k,\mu)$ coherently enhances the observed spectrum which makes similar size of but opposite contribution to the redshift space power spectrum as $T(k,\mu)$. So the combined effect of $F(k,\mu)$ and $T(k,\mu)$ on the observed spectrum can be smaller. But for the completeness of the order expansion, we include both contributions in our analysis.

\section{finding the fog function}
\label{sec:fog}

If the full indefinite higher order polynomials in Eq.~(\ref{eq:Pkred_final}) are given, the residual one--point FoG, $D^{\rm FoG}$, appears as a complete function of the single parameter $\sigma_z$ which is independent of scale \cite{Zheng13}. But, in reality, the perturbative terms in Eq.~(\ref{eq:Pkred_final}) are incompletely known, the assumed scale independence will not be available at the specific scale. In this work, the higher order polynomials contain the order contributions of density--velocity cross--correlations and velocity--velocity auto--correlations up to $j_1^2$ order. The cutoff scale is placed at the regime in which those low order polynomials are dominating. Within this limit, $D^{\rm FoG}$ is described by a scale independent coherent curve which can be pre--determined in cosmological model independent way.

\subsection{The best fitted FoG function}
\label{subsec:bestfog}
The residual FoG term is, by definition, independent of the separation vector $\bf x$ between two points, and factored out of the integration in terms of $\bf x$. Then the remaining integrand contains all the correlation pairs, and the external multiplicity factor $D^{\rm FoG}$ depends on only one--point random velocity dispersion and its cumulants. This one--point FoG factor is described by a function of $k\mu$ in Fourier space. If we fully divide the perturbative terms $P_{\rm perturbed}(k,\mu)$ from the observed spectra $P^{\rm (S)}_{\rm sim}(k,\mu)$ from simulations, the residual multiplicity factor will be consistent with the true $D^{\rm FoG}$, 
\beq\label{eq:fogfig}
D^{\rm FoG}=\frac{P^{\rm (S)}_{\rm sim}(k,\mu)}{P_{\rm perturbed}(k,\mu)},
\eeq
which should be described by a single function in terms of $k\mu$ regardless of the individual $k$ modes. We present the measured residual FoG term after dividing out different perturbative term combinations in Fig.~\ref{fig:FoGres}.

\begin{figure*}
\centering
\includegraphics[width=1.0\columnwidth]{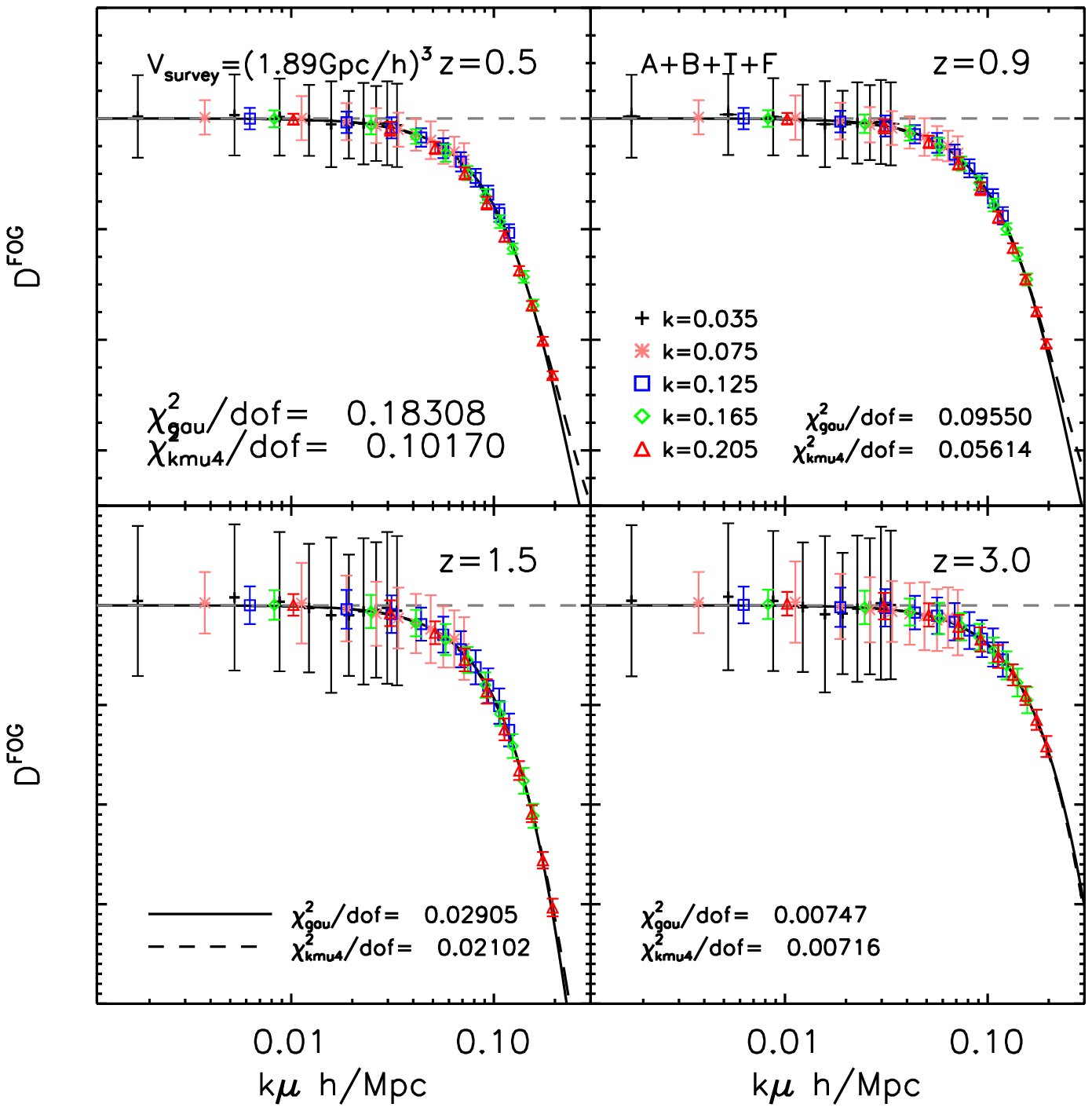}
\includegraphics[width=1.0\columnwidth]{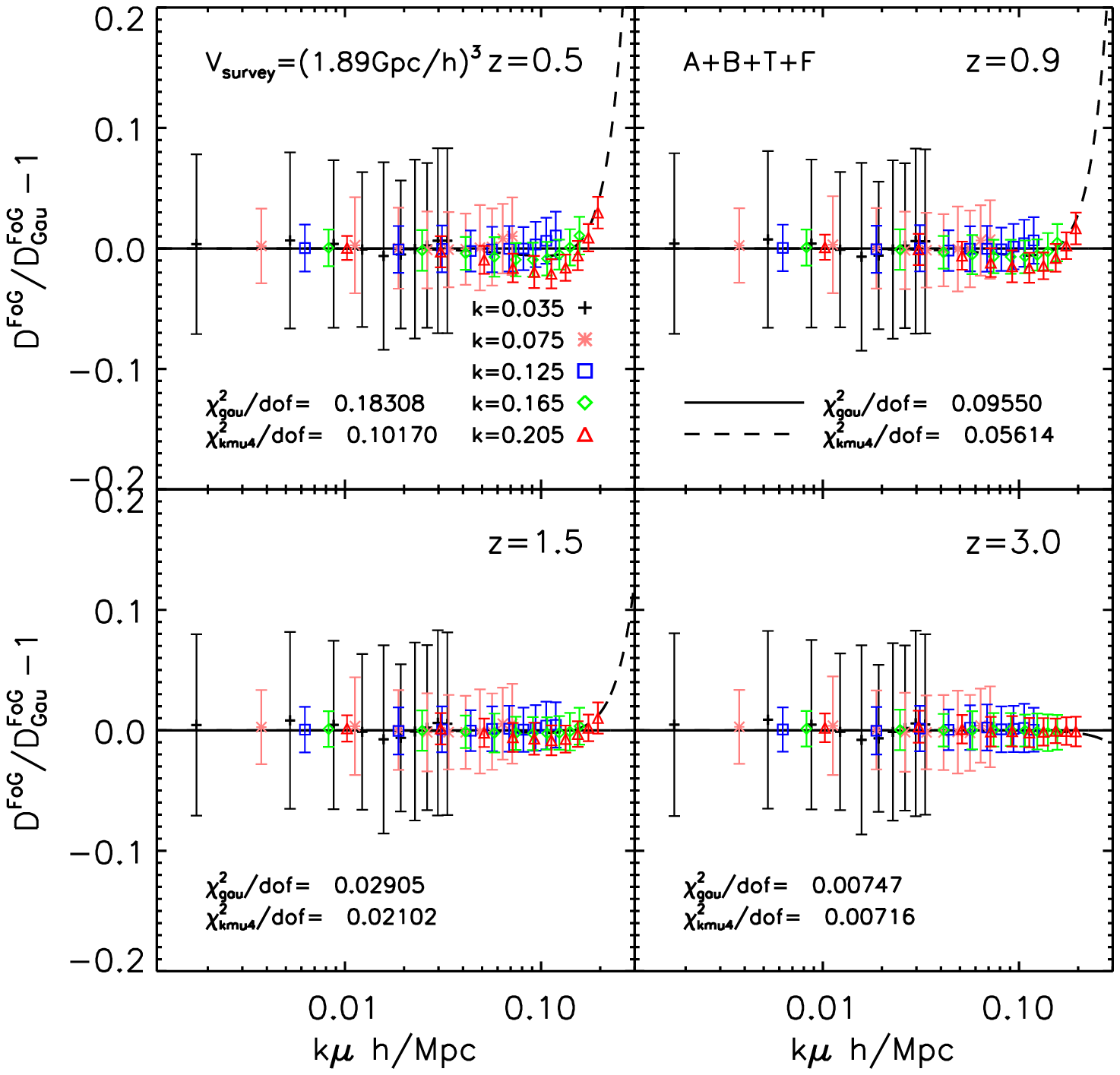}
\caption{\label{fig:dFoG} {\em Left panel:} The measured FoG is presented at $z=0.5$, $0.9$, $1.5$ and $3.0$, with errors coming from the sample standard deviation of $P^{\rm (S)}_{\rm sim}(k,\mu)$. The measured points are central values from 100 realisation simulations. The solid and dashed lines are fitted FoG terms using data at $k\sim0.035-0.205\hompc$. {\em Right panel:} The difference between the measured and best fit Gaussian FoG terms. }
\end{figure*}

To begin with, the residual FoG is presented after dividing out only $j_1^0$ order terms at $z=0.5$ in the first column of Fig.~\ref{fig:FoGres}. It is corresponding to Scoccimarro's model in which non--linear corrections are complete for all zero's order terms, but no higher polynomials are added \cite{Scoccimarro04}. The residuals are computed using different scales of $k=(0.065, 0.095, 0.135, 0.175, 0.205)\hompc$. Those residuals at different $k$ modes are not aligned coherently along a single curve. It indicates that the measured residuals are contaminated by other perturbative terms which are not ignorable at those scales.

The improved RSD theoretical model, TNS model \cite{Taruya10}, has been suggested to reproduce BAO features in precision . This standard perturbation RSD theory additionally includes one $j_1^{1}$ and one $j_1^{2}$ order terms, which are $A(k,\mu)$ and $B(k,\mu)$ respectively. Other contributions in $j_1^{2}$ order were not included, as those are considered to be irrelevant for BAO reconstruction. The measured residuals are presented in the second column of Fig.~\ref{fig:FoGres}, which exhibits more consistent $D^{\rm FoG}$ at different scales, although there is still slight inconsistency at high $k$ modes and large $\mu$ bins. Here the uncertainty due to non--linear corrections is removed by directly measuring both $A$ and $B$ terms, and the result can be improved by completing all polynomials at the given order.

Next we present new result after dividing out $j_1^{1}$ and $j_1^{2}$ order polynomials generated by all density--velocity cross--correlations, i.e. including $T(k,\mu)$ as well. The $T(k,\mu)$ based on the tri--point correlation function has not been considered seriously yet, as it is assumed not to be significant for BAO reconstruction, and the full non--linear corrections are not known. In this manuscript we are interested in full RSD clustering analysis more than just focusing on distance measures, and we are able to calculate non--linear corrections using simulations. The residuals are presented in the third column of Fig.~\ref{fig:FoGres}. The measured FoG is aligned more consistently along a single curve which is different from the previous case of $A+B$ division. It is interesting to note that the effect of $T$ is not negligible at all. If the information of full RSD clustering is requested, then $T$ should be counted. Finally, we also include another $j_1^{2}$ order term originated from velocity--velocity auto correlation higher order polynomials, $F(k,\mu)$. All measured residuals at different scales are exhibited consistently up to $k\la0.2h/$Mpc presented in the fourth column of Fig.~\ref{fig:FoGres}. From now on, the residual FoG is derived after dividing out all $A$, $B$, $T$ and $F$ terms. The fractional differences are presented in the bottom panel of Fig.~\ref{fig:FoGres}. The measured residual FoG's of $A+B$ and $A+B+F+T$ divisions are ignorable for the low resolution experiment targeting 5\% observables, but can be significant for the high resolution experiment targeting 1 or 2\% observables. The consistency in terms of $k\mu$ is slightly better with $A+B+T$ than with $A+B+F+T$, but we are not able to justify this in terms of the full perturbation theory presented in the manuscript. It can be accidental improvement, or there might be deeper physical implications, which is not yet known. We prove that we need higher order contribution of $T$, but we need to study further to fully understand the result.

The detailed test on the residual FoG is provided in Fig.~\ref{fig:dFoG}  to determine the best FoG functional form for the resolution of $(1.89\, {\rm Gpc}/h)^3$ box which is close to the survey volume of  most future experiments such as DESI. First of all, errors on the residual FoG are estimated in the following way. Averages of all perturbative terms at $j_1^{0}$, $j_1^{1}$ and $j_1^{2}$ are computed using 100 realisation simulations. The observed spectrum for each box is divided by this averaged perturbative templates, which provides the dispersion of the measured residuals as presented in the left panel of Fig.~\ref{fig:dFoG}. Then we fit the FoG model in Eq.~\ref{eq:fog_phenom}. The $\sigma_z$ is assumed to be constantly variable for all FoG models. The $k$ space is binned from $k_{\rm min}=0.035\hompc$ to $k_{\rm max}=0.205\hompc$ with $k$ spacing of $\Delta k=0.01\hompc$. 
The solid curves in the left panel of Fig.~\ref{fig:dFoG} represent the best fitted Gaussian FoG term at the given $k_{\rm cut}=0.205\hompc$. Even though $E_{2n}$ is set to be zero, this simplest Gaussian FoG model fit to the measured residual FoG up to $k=0.2\hompc$. The dashed curves represent the result when both $\sigma_z$ and $E_{2n}$ are fitted. We practicably only consider $E_4$ in our test, but the results are representative of the general $E_{2n}$ case. In the right panel of Fig.~\ref{fig:dFoG}, we present the difference between the measured and estimated FoG using Gaussian model. Considered the resolution of map with $(1.89\, {\rm Gpc}/h)^3$ volume, the estimated FoG model up to $k\simeq 0.2\hompc$ is reasonably placed within errors.

\subsection{The fitted $\sigma_z^2-k$ relation}

\begin{figure}
\centering
\includegraphics[width=1.0\columnwidth]{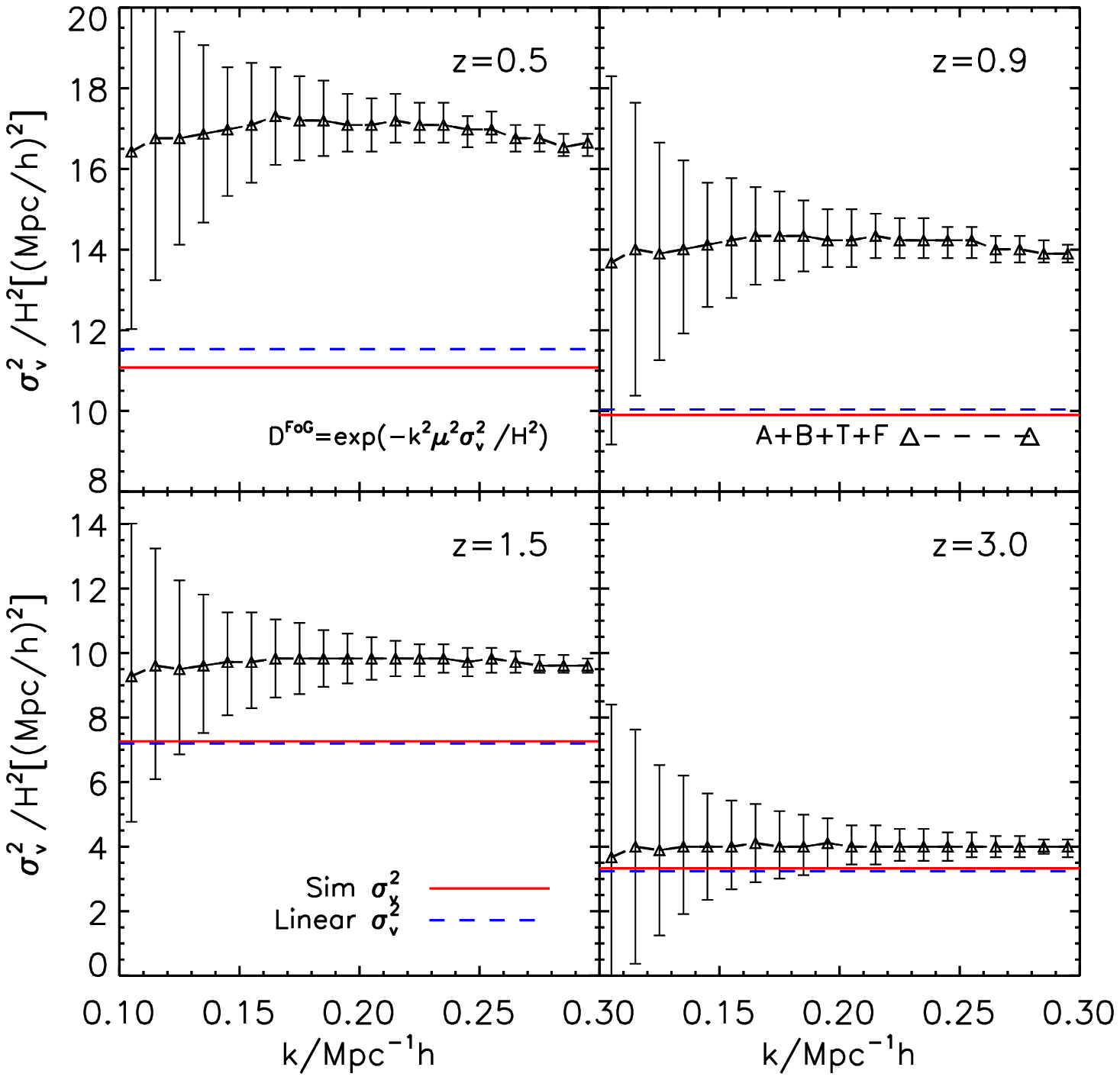}
\caption{The FoG is assumed to be Gaussian, and we fit best $\sigma_v$ for the measured residuals at each $k$, to test whether the observed $\sigma_v$ is nearly scale dependent or not. The measured residuals are presented as triangle points, and errors are explained in the text. The solid and dash lines separately represent the simulation measured and linear one dimensional velocity dispersions.}
\label{fig:sigz_ABFT}
\end{figure}

Note that we have to be assured that whether there is any scale dependence on the measured $\sigma_z$. In Fig.~\ref{fig:sigz_ABFT}, we present the fitted $\sigma_z$ in each $k$ mode, which contains 10 equally distributed $\mu$ bins from $\mu=0$ to 1,  at $z=0.5$, 0.9, 1.5 and 3.0 using Gaussian FoG model. The observed spectra are minimally contaminated by non--linear physics at high redshift, e.g. $z=3.0$. The observed $\sigma_z$ at $z=3.0$ is indeed constant regardless of $k$ scale up to $k\la 0.3\hompc$, when the full $A+B+T+F$ terms are used for the division. Here we do not present results at $k\la 0.1\hompc$, because the measured residual is nearly featureless to determined $\sigma_z$. There will be no planned redshift survey at this high redshift, but we confirm our methodology at the region in which the measured spectrum remains nearly linear. We continue our test at lower redshifts of $z=0.5$, $0.9$ and $1.5$ in which the observed spectra are highly contaminated by non--linear physics. There is slight $k$ dependence of $\sigma_z$ developed between $0.1\la k \la 0.15$, but the variation of $\sigma_z$ is much smaller than errors which are expected from the future survey. If the full $A+B+T+F$ terms are not used for the division, the measured $\sigma_z$ becomes scale--dependent at $k \ga 0.15\hompc$. The assumption of coherent $\sigma_z$ is valid only if $A+B+T+F$ division case.

\bfib{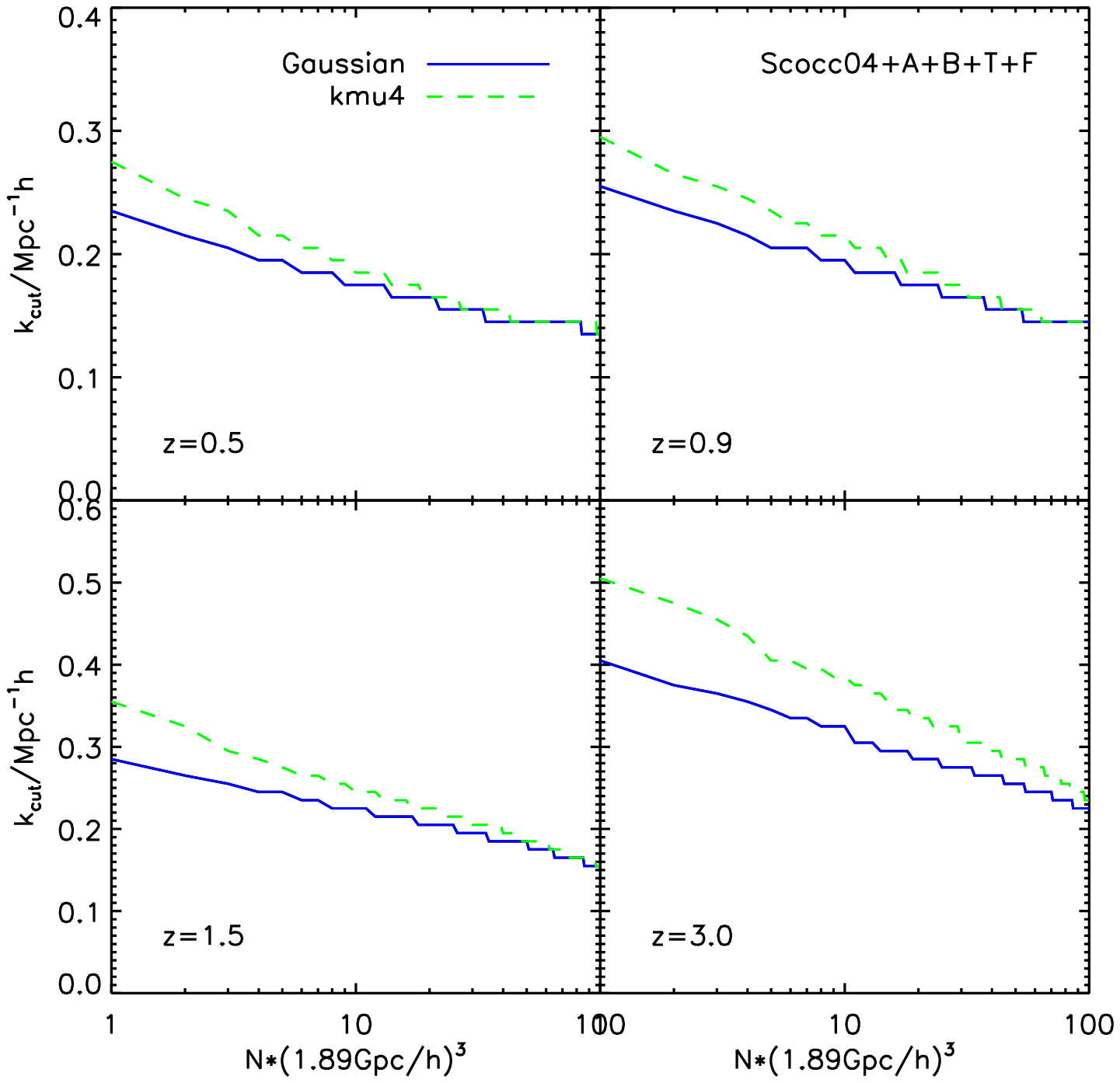}
\caption{The fitted $k_{\rm cut}-V_{\rm survey}$ relation for $A+B+T+F$ model. In general, Gaussian FoG function is better than other fitting functions. This figure could be a guideline of the maximum $k$ mode we can use for a specific survey strategy in the future.}
\label{fig:kcut}
\efib

In Fig.~\ref{fig:sigz_ABFT}, it is interesting to note the difference between the fitted $\sigma_z^2$ and the directly measured $\sigma_z^2$ from simulations. As mentioned in \cite{Zhangrsd,Zheng13}, the one--point FoG part could be well approximated by a Gaussian function with the measured $\sigma_z^2$ as input. The accuracy is within $1\%$ level at $k<0.3\hompc$. The difference here is possibly caused by the contamination due to the higher order polynomials than $j_1^2$, or by the unknown non--linear dynamics of random velocity fields. We confirm that the coherent $\sigma_z$ treatment works fine to fit the measured RSD spectrum, but the detailed theoretical mechanism to compute $\sigma_z$ remains to be unknown, which is our next subject to be explored.



\bfig{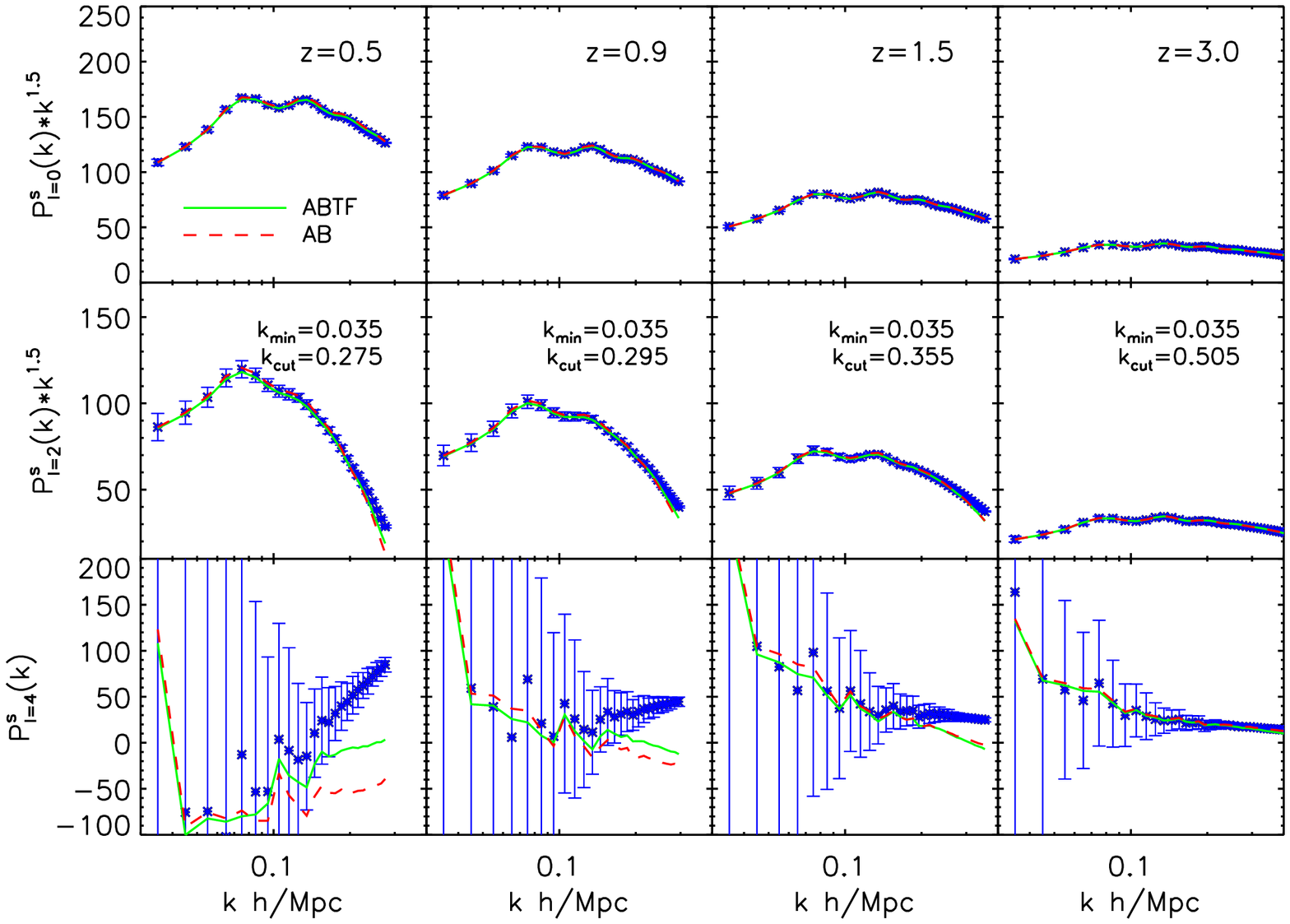}
\caption{The multipole components of spectrum is presented at $z=0.5$, $0.9$, $1.5$ and $3.0$ from the first to fourth columns. Monopole, quadrupole and hexadecapole spectra are shown in the top, middle and bottom panels. The green solid curves represent the theoretical estimation using $A+B+T+F$ combination, the red dash curves represent the theoretical estimation using $A+B$ combination, and the cross points represent the directly observed multipoles. The $\sigma_z$ is the best fit value for the correspondent model. The error bars are the sample standard deviation from 100 simulation realizations.}
\label{fig:pkl} 
\efig

\begin{figure}
\centering
\includegraphics[width=1.0\columnwidth]{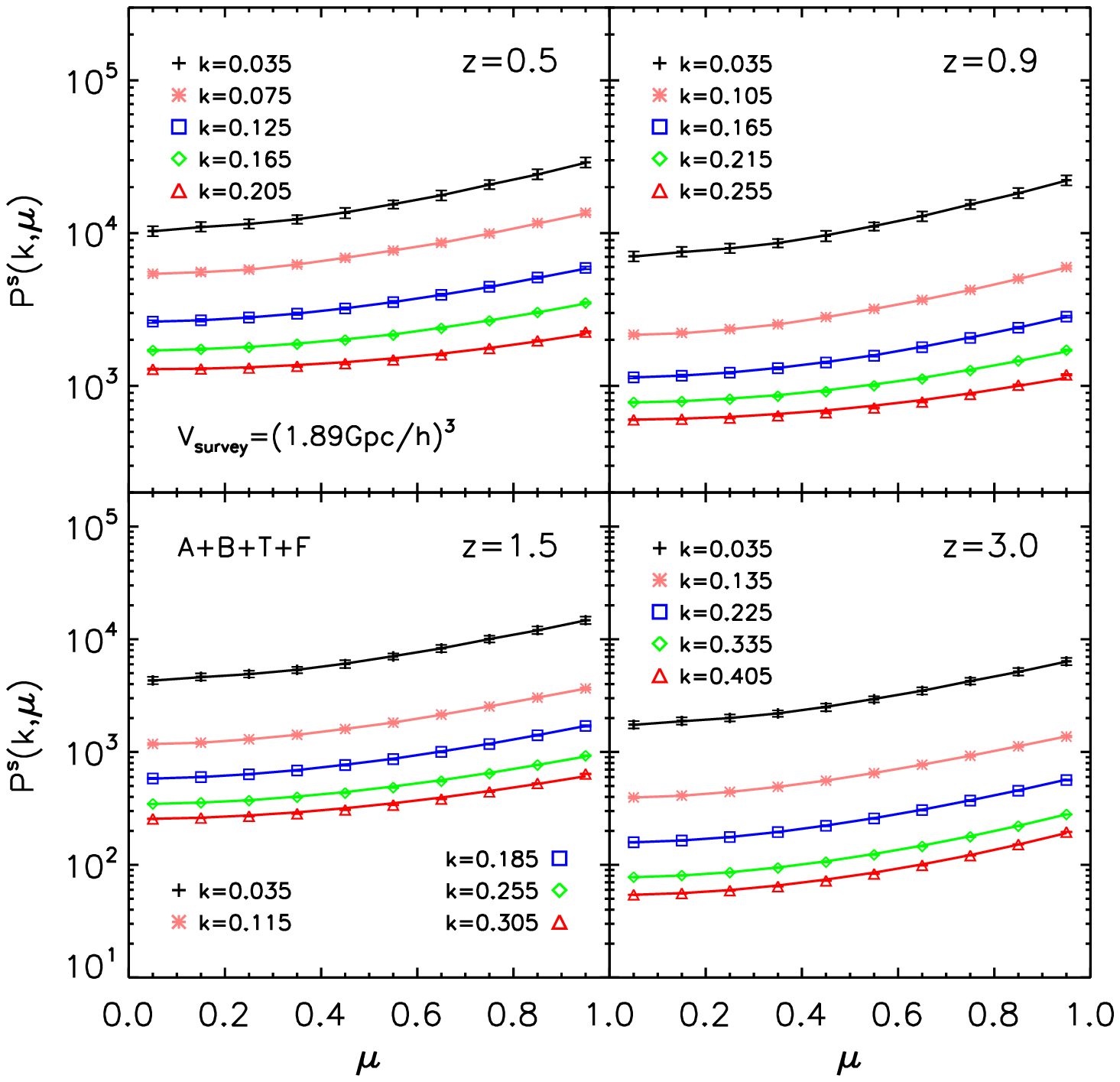}
\caption{The 2D anisotropy spectra are presented at $z=0.5$, $0.9$, $1.5$ and $3.0$. The points with errors represent the direct measurements from the simulations, and the solid curves represent the theoretical estimation using our method. The error bars are the sample standard deviation from 100 simulation realizations.}
\label{fig:pkmu}
\end{figure}

\section{discussions}
\label{sec:discussion}

We study the non--linear mapping of dark matter clustering from real space to redshift space. The perturbative description with one--point FoG prior is proved to be a precise RSD theoretical model predicting the observed spectrum in the redshift space at $k\la 0.2\hompc$. However, this perturbative theory is incomplete, and the model contains an unknown sub--dominant contamination even at $k\la 0.2\hompc$ due to higher order terms which are not included in the finite expansion. In Fig. \ref{fig:kcut}, we define $k_{\rm cut}$ as the scale at which the systematic uncertainty due to this contamination is smaller than the cosmic variance of the survey volume. That is the $k_{\rm max}$ at which the reduced $\chi^2$ becomes greater than 1, $\chi^2_{\rm reduced}\ga 1$. The conservative bound will be imposed, if we are concerned with bigger survey volumes, or if we wishes to control systematics much smaller than the survey variance. In Fig.~\ref{fig:kcut}, we present bounds of $k_{\rm cut}$ according to the survey volumes represented by $N\times\,$survey volume (which we defined as our simulation volume, then the survey variance $\propto 1/\sqrt{N}$).  When $N=10$, the contamination is controlled to be smaller approximately by a factor of 3 than the cosmic variance of one survey volume. Note that there is little change in $k_{\rm cut}$ until $N\sim 10$. If it is requested under 10\% uncertainty control, then $k_{\rm cut}$ is reduced by $0.1\hompc$ at all redshifts.



We estimate the RSD power spectrum using the measured perturbative templates of $A(k,\mu)$, $B(k,\mu)$, $T(k,\mu)$, $F(k,\mu)$, and the predetermined Gaussian FoG function with a variable constant $\sigma_z$, from simulations. The templates are computed using averages of 100 realisations of which variation is as small as 10\% of statistical uncertainty of each simulation. We ignore this small variation in predicting the RSD power spectrum using these templates. Higher order polynomials of $j_1^n$ with $n\ge 3$ are not included in our RSD model, as we remove modes at $k\ga 0.2\hompc$ in which those higher order terms begin to be dominant. In our treatment, the systematic due to unknown non--linear corrections is resolved as those are exact measurements from the simulations. Also, the systematic due to the randomness of peculiar velocities of particles is controlled by the predetermined Gaussian FoG function, which is tested to be the best option not only theoretically but also observationally. The $\sigma_z$ remains undetermined, but proved to be constant at $k\la 0.3\hompc$. When the best fitting $\sigma_z$ is known, the theoretical RSD spectrum is provided. However, non--Gaussianity FoG effect with non--zero $E_{2n}$ develops non--negligibly at $k\ga 0.2\hompc$. 

We present the multipole components of the estimated RSD spectrum in Fig.~\ref{fig:pkl}. Monopole moments are presented in the top panels at $z=0.5$, $0.9$, $1.5$ and $3.0$ from the first to the fourth column. The cross points represent the direct monopole measurement from the simulations, and the solid green curves represent the estimated RSD spectrum described in the previous paragraph. The measured and estimated spectra are consistent to each other. Quadrupole moments are more contaminated by all systematics caused by non--linear clustering mapping from real to redshift spaces, which are presented in the middle panels. The measured ones agree with the estimation (green lines) in precision within $2\%$ by  $k=0.185\hompc$ at $z=0.5$, $k=0.225\hompc$ at $z=0.9$, $k=0.265\hompc$ at $z=1.5$, and $k=0.395\hompc$ at $z=3.0$. It is a remarkable improvement in our model. In the bottom panels, we compare hexadecapole moments as well. Although there are discrepancies observed at low redshifts, the measured errors are big there as well. The measured and estimated hexadecapole moments are still consistent under the error budget. The red dash curves represent the theoretical estimation using $A+B$ combination. The difference is ignorable with low resolution experiments, but significant with high resolution future experiments.

Alternatively, the RSD spectrum can be presented in the 2D anisotropy plane of $k$ and $\mu$, as denoted by $P^s(k,\mu)$. We select spectra at five different scales between $k_{\rm min}$ and $k_{\rm cut}$ at each redshift bin, and present those in $\mu$ direction. The $P^s(k,\mu)$ at $\mu\rightarrow 1$ are contaminated most by RSD systematics. The directly measured $P^s(k,\mu)$ are presented as cross points in Fig~\ref{fig:pkmu}, and our estimated $P^s(k,\mu)$ are presented as solid curves. Both are consistent in great precision even at $\mu\rightarrow 1$ limit. In the following works, we will investigate the detailed power spectra reconstruction, the mapping of halo or galaxy clustering, and the effect of higher than $j_1^2$ order polynomials.

We work out the detailed test of RSD perturbation theoretical models. However, the measured templates will be directly exploited to RSD anisotropy analysis as well. In our following work, we investigate the growth function dependence on the measured templates, and feed the outputs into data analysis, without invoking to theoretical RSD modelling.

\section*{Acknowledgments}

We thank Pengjie Zhang, Atsushi Taruya and Shun Saito for useful discussions and thank Minji Oh for providing the N-body simulation data. We thank the anonymous referee for helpful comments on the connectivity of $T$ term and its numerical accuracy. The work of running simulation was supported by the National Institute of Supercomputing and Network/Korea Institute of Science and Technology Information with supercomputing resources including technical support (KSC-2015-C1-017). Numerical calculations were performed by using a high performance computing cluster in the Korea Astronomy and Space Science Institute. 






\appendix
\section{The measured $T(k,\mu)$ term}
\label{app}

\begin{figure*}
\centering
\includegraphics[width=1.0\columnwidth]{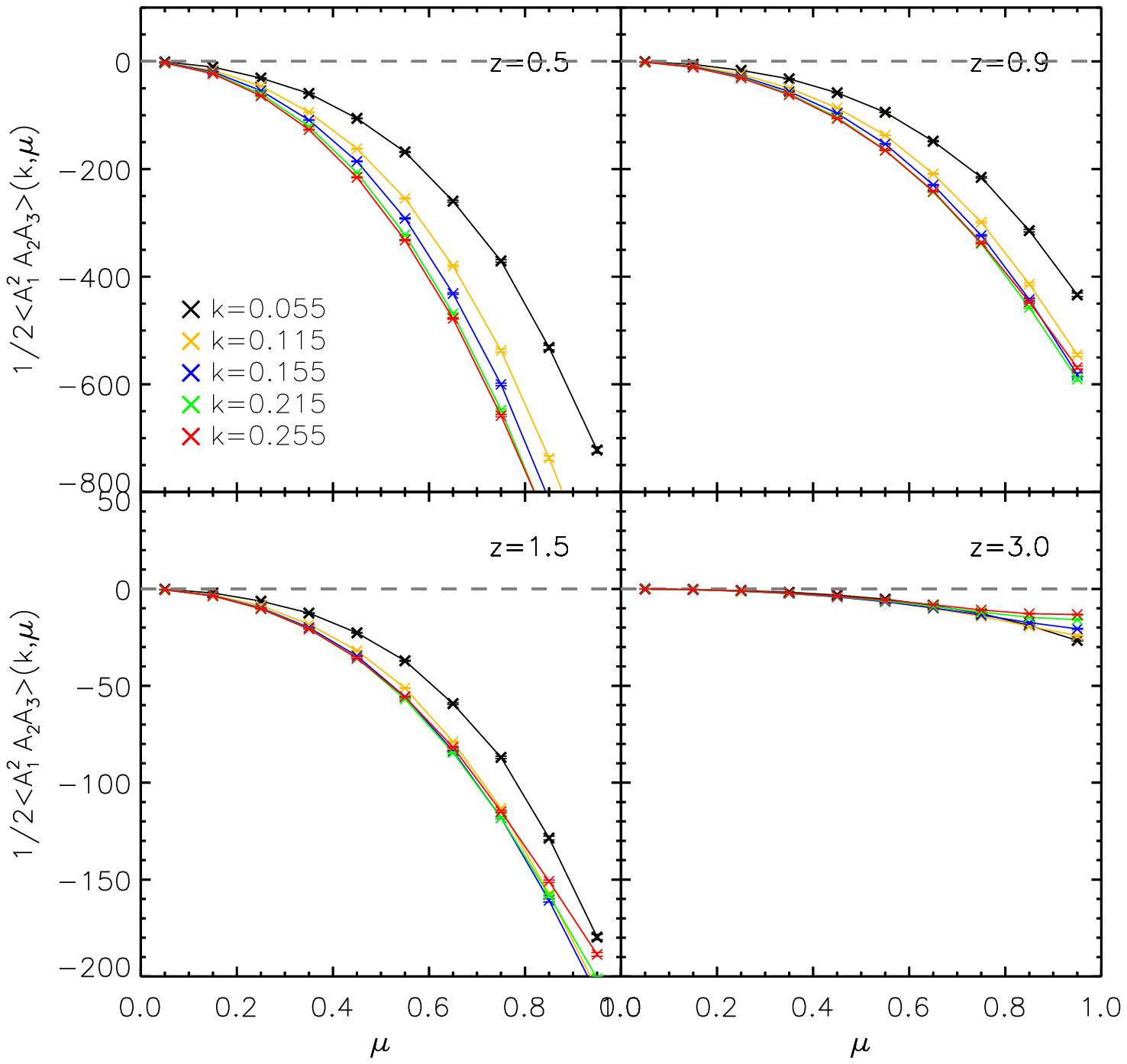}
\includegraphics[width=1.0\columnwidth]{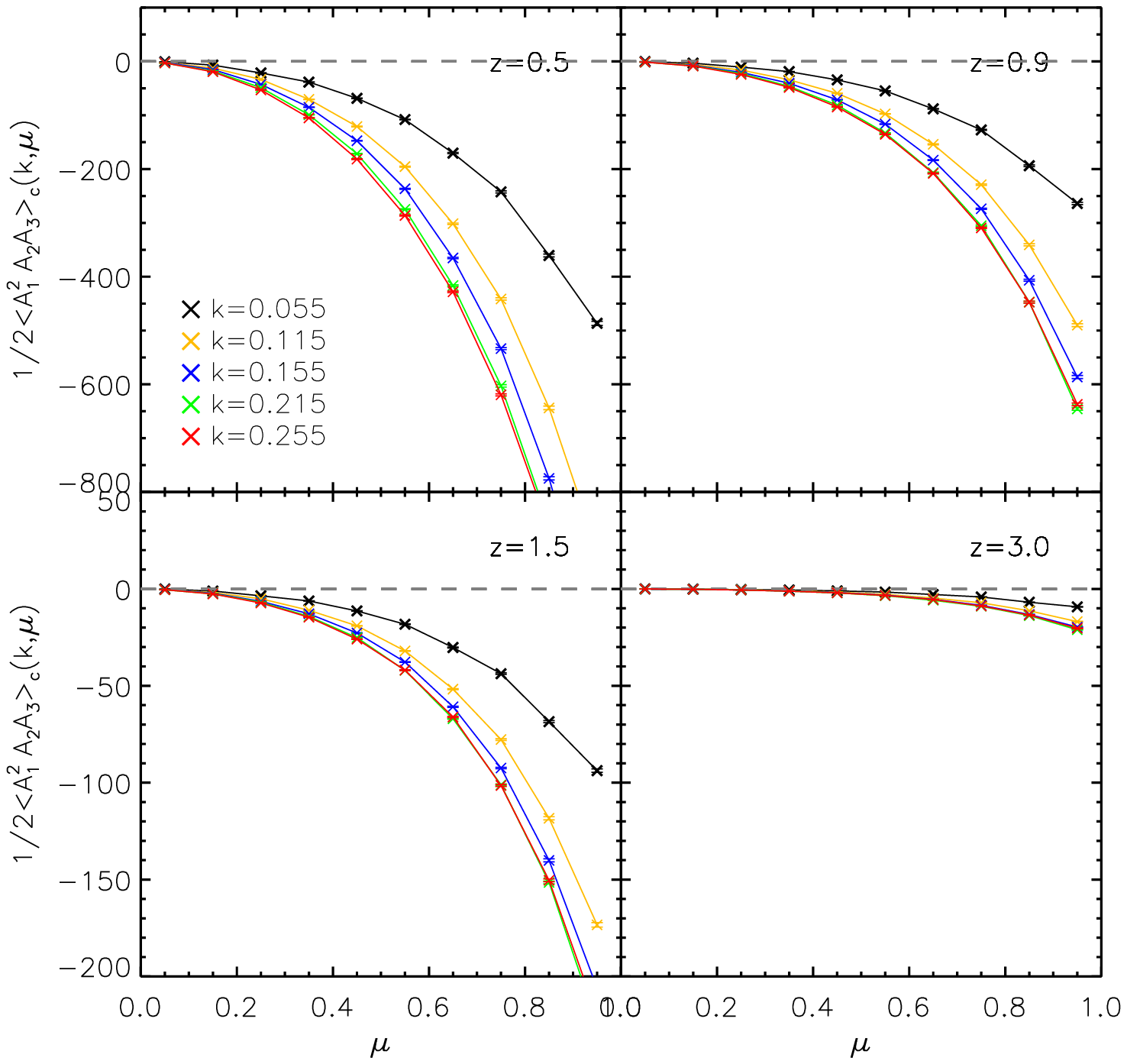}\\
\includegraphics[width=1.0\columnwidth]{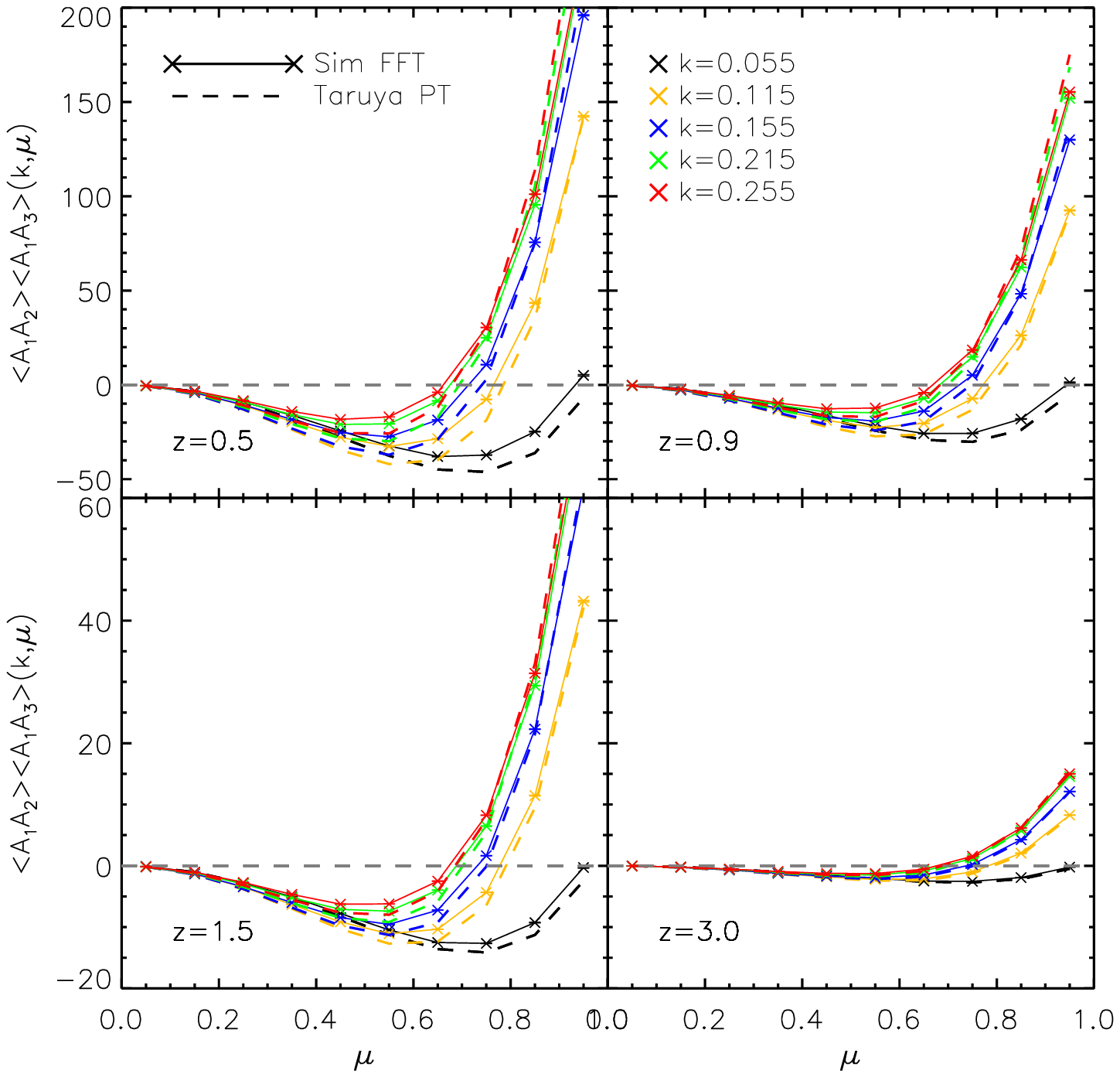}
\includegraphics[width=1.0\columnwidth]{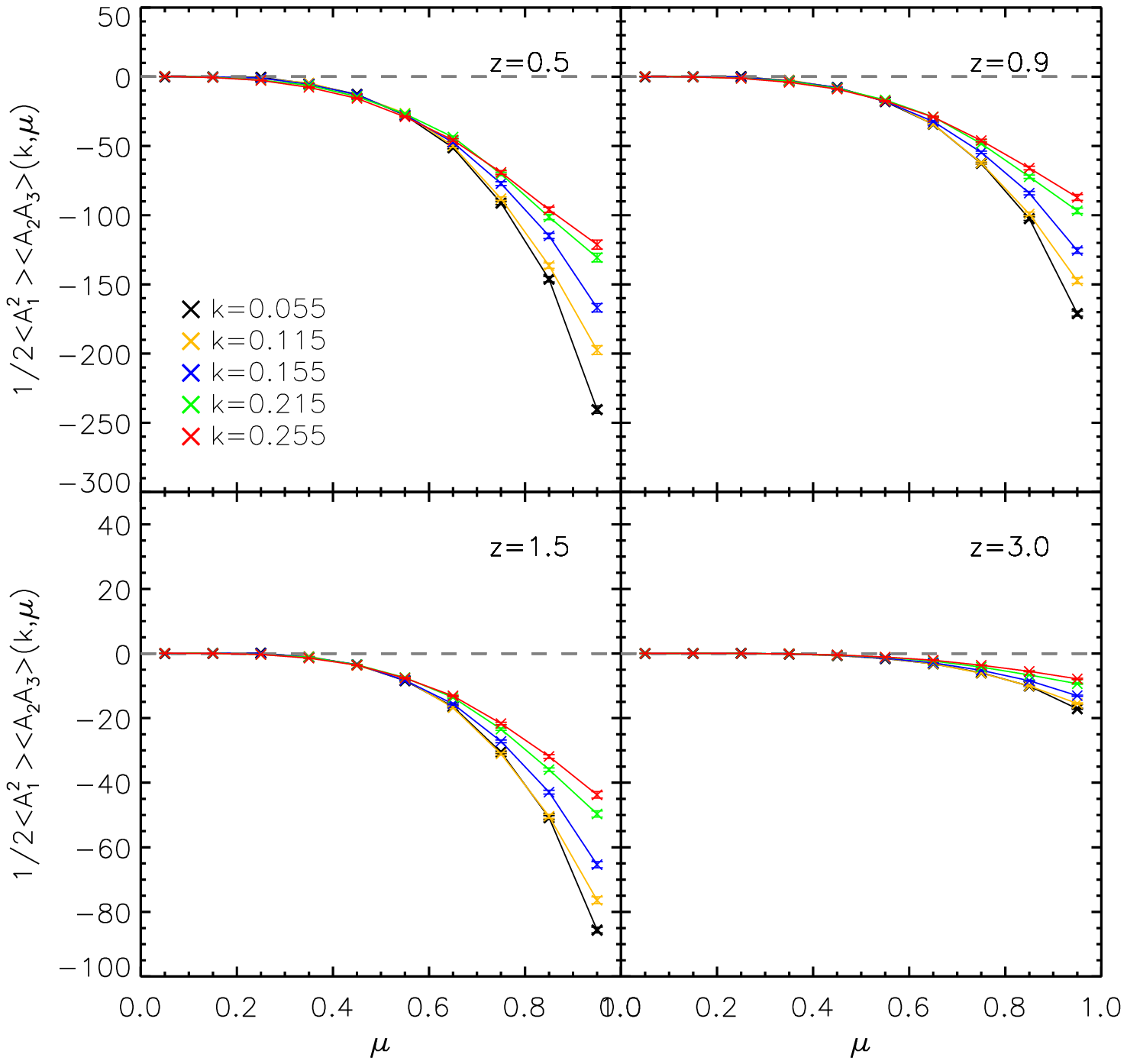}
\caption{The detailed procedure to compute $T(k,\mu)$ is presented. The measured full $\langle A_1^2A_2A_3\rangle/2$ and the connected $T(k,\mu)$ measurement are presented in the top left and right panels respectively. The unconnected pairs of $\langle A_1A_2\rangle\langle A_1A_3\rangle$ and $\langle A_1^2\rangle\langle A_2A_3\rangle/2$ are measured in the bottom panels.}
\label{fig:app1}
\end{figure*}

\begin{figure}
\centering
\includegraphics[width=1.0\columnwidth]{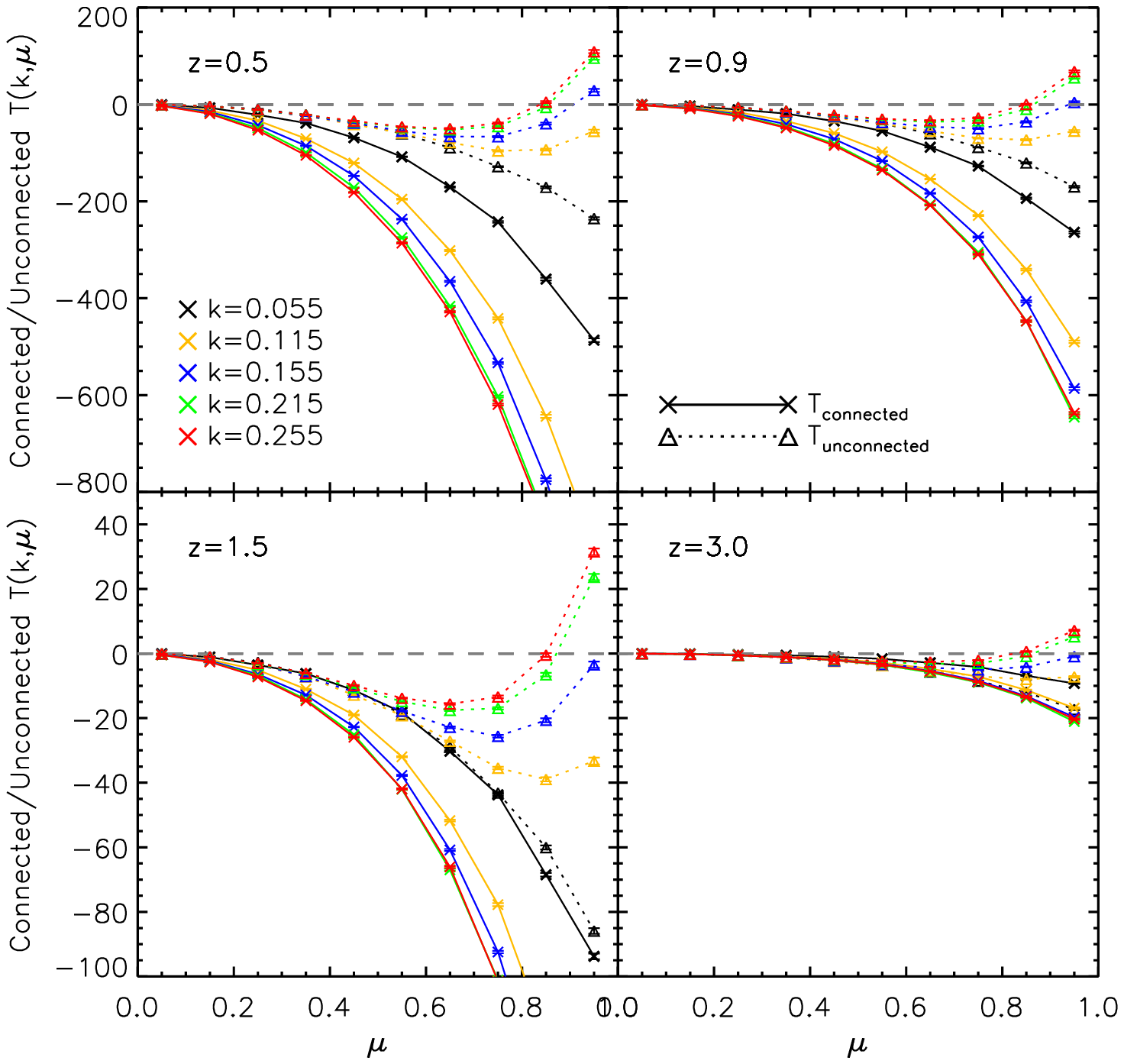}
\caption{We present the connected $T(k,\mu)$ as solid curves, and the unconnected $T(k,\mu)$ as dotted curves.}
\label{fig:app3}
\end{figure}

We present the measured $T(k,\mu)$ in this Appendix section. The connected $T(k,\mu)$
\bea
T(k,\mu)&=& \frac{1}{2} j_1^2\,\int d^3\bfx \,\,e^{i\bfk\cdot\bfx}\,\,\langle A_1^2A_2A_3\rangle_c \nonumber \\
&=& \frac{1}{2} j_1^2\,\int d^3\bfx \,\,e^{i\bfk\cdot\bfx}\,\,\left\{\langle A_1^2A_2A_3\rangle\right. \nonumber\\
&&\left.-\langle A_1^2\rangle\langle A_2A_3\rangle-2\langle A_1A_2\rangle\langle A_1A_3\rangle\right\}
\eea
 is computed by subtracting the unconnected pairs from $\langle A_1^2A_2A_3\rangle/2$	. The full $T$ term, $\langle A_1^2A_2A_3\rangle/2$, is presented in the top left panel of Fig.~\ref{fig:app1}. There is no subtraction, and we integrate all effects here. The unconnected pair multiplications are presented in the bottom panels of Fig.~\ref{fig:app1}. Both measured and theoretically estimated $\langle A_1A_2\rangle\langle A_1A_3\rangle$ ($B$ term) are presented as solid and dash curves in the bottom left panel respectively, which shows a good consistency. The measured $\langle A_1^2\rangle\langle A_2A_3\rangle/2$ is presented in the bottom right panel. The resultant connected $T(k,\mu)$ is computed by subtracting those unconnected pairs from $\langle A_1^2A_2A_3\rangle/2$, which is shown in the top right panel. In Fig.~\ref{fig:app3}, the difference between the connected and unconnected $T(k,\mu)$ is presented. The solid curves present the connected $T(k,\mu)$, and the dotted curves present the unconnected $T(k,\mu)$. As it is expected, the connected part gets gradually more prominent relative to the unconnected part at lower redshfit and at higher $k$ modes.

\begin{figure}
\centering
\includegraphics[width=1.0\columnwidth]{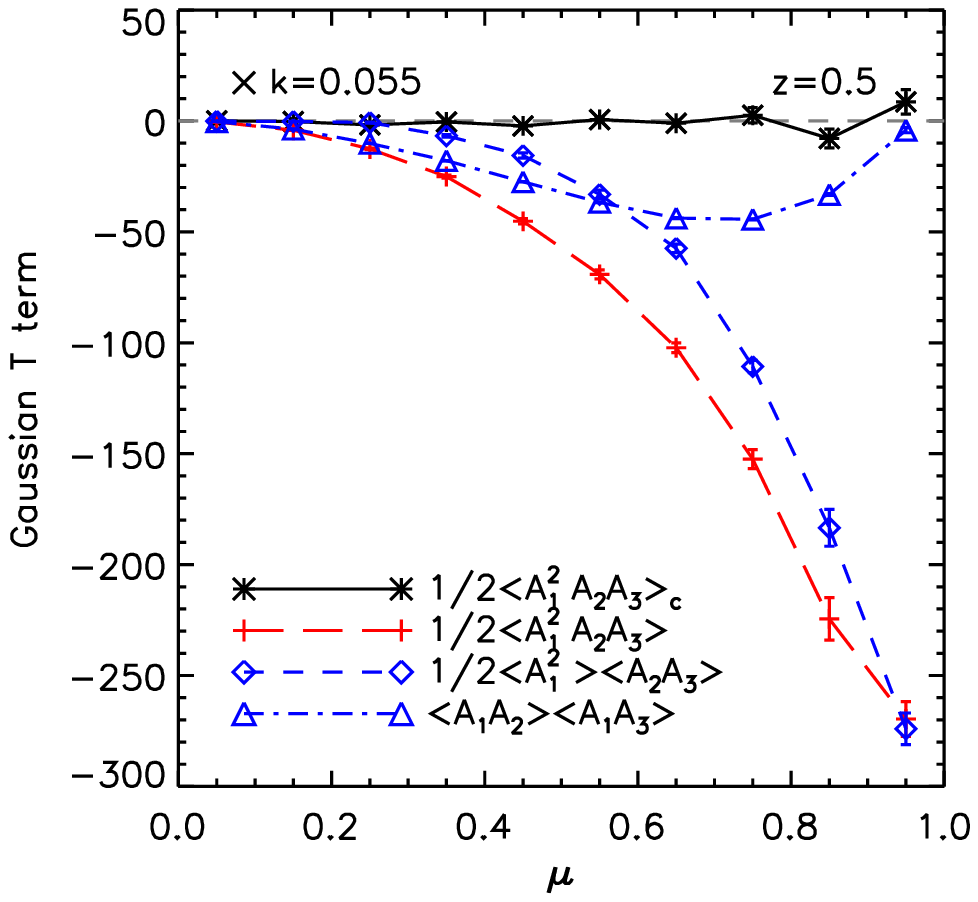}
\caption{We present the connected $T(k,\mu)$ (black solid curve) using the Gaissian density field at $z=0.5$ and $k=0.055\ompc$. Red long dash curve represents the full $\langle A_1^2A_2A_3\rangle/2$, and blue dash--dotted and blue dash curves represent $\langle A_1A_2\rangle\langle A_1A_3\rangle$ and $\langle A_1^2\rangle\langle A_2A_3\rangle/2$ respectively.}
\label{fig:app2}
\end{figure}

We test the validity of our $T$ term measurement using the Gaussian field simulation. The initial condition of simulation is defined at $z=49$ in which the distribution of matters is approximately Gaussian. The density and velocity fields are enhanced by the linear growth factor/rate ratios between $z=0.5$ and $z=49$. Since the field is Gaussian, the connected $T(k,\mu)$ is expected to be ignorable, as shown by the theoretical formulation of $T(k,\mu)$ derived at the lowest order in~\cite{Taruya13}. We take $k=0.055h$/Mpc for example, in Fig~\ref{fig:app2}, the full $\langle A_1^2A_2A_3\rangle/2$ is presented by a red long dash curve. The unconnected parts $\langle A_1^2\rangle\langle A_2A_3\rangle/2$ and $\langle A_1A_2\rangle \langle A_1A_3\rangle$ are presented by blue dash and blue dot-dashed curves respectively. When both are subtracted from the full $\langle A_1^2A_2A_3\rangle/2$, the connected $T(k,\mu)$ is measured to be nearly zero, which is expected from the theory. 

We calculate the same results at higher $k$ modes and also find them very close to 0. This gives us confidence about our calculation of $T$ term in this work, and it would be an interesting issue to discuss that why (tree-level) perturbation theory failed to describe $T$ term even at very large scales. Further investigations will be provided in our future work.

\end{document}